\documentclass[particles,perspective,submit,pdftex,moreauthors]{Definitions/mdpi} 
\firstpage{1} 
\pubvolume{1}
\issuenum{1}
\articlenumber{0}
\pubyear{2026}
\copyrightyear{2026}
\datereceived{ } 
\daterevised{ } 
\dateaccepted{ } 
\datepublished{ } 
\pdfoutput=1 

\preto{\abstractkeywords}{\nolinenumbers}

\Title{Kinetic Simulations of Laser-Driven Compression and Heating of Magnetised Cryogenic Hydrogen Targets using PIConGPU}

\Author{
Filip Opto{\l}owicz$^{1,*}$\orcidA{},
Klaus Steiniger$^{2,3}$\orcidK{},
David Blaschke$^{1,2,3}$\orcidB{},
Michael Bussmann$^{2,3}$\orcidC{} and
Brian Marre$^{2}$\orcidM{}
}

\AuthorNames{Filip Opto{\l}owicz, Klaus Steiniger, David Blaschke,  Michael Bussmann, Brian Marre
}

\address{%
$^{1}$ \quad Institute of Theoretical Physics, University of Wroclaw, Max Born Pl. 9, 50-204 Wroclaw, Poland\\
$^{2}$ \quad Helmholtz-Zentrum Dresden-Rossendorf (HZDR), Bautzner Landstrasse 400, 01328 Dresden, Germany\\
$^{3}$ \quad Center for Advanced Systems Understanding (CASUS), Untermarkt 20, 02826 G\"orlitz, Germany\\
}

\corres{Correspondence: filip.optolowicz@uwr.edu.pl
}

\abstract{
We present fully kinetic two-dimensional, three-velocity-component (2D3V) PIConGPU simulations of a three-beam direct-drive interaction with a 15~$\mu$m solid-density cryogenic hydrogen cylinder, establishing a predictive numerical baseline for the operational DRACO ($\tau = 30$~fs) and upcoming PEnELOPE ($\tau = 150$~fs) laser facilities at HZDR \cite{DRACO_Publ-Id:43136/1}. The simulations resolve charge-separation fields on the order of 3~TV/m and reveal a robust kinematic bifurcation of the accelerated population into a fast (1--5~MeV) ion beam and a slower bulk (1--100~keV) flow. We demonstrate analytically and numerically that the charge-separation front ($v_{hb}$) is an intrinsically non-quasi-neutral electrostatic double layer that lies outside the closure assumptions of radiation-hydrodynamic models. A simple $2v_{hb}$ reflection scaling derived directly from the front trajectory tracks the centroid of the constant-energy fast-ion band under the impulsive 30~fs driver and the time-varying upper edge of the swept fast-ion band under the sustained 150~fs driver, across both intensities ($a_0 = 12.7$ and $22.0$), establishing this non-thermal mechanism as the dominant acceleration pathway. We then scan an external axial magnetic field from 0~T to 10~kT. Laboratory-achievable 20~T fields leave all macroscopic observables unchanged; fields at the kT scale progressively magnetise the MeV hot-electron population, quench the laser-driven charge-separation mechanism, suppress the fast-ion band, and more than double the net-inward compression time of the short-pulse driver---while extending the outer target envelope. A geometric equivalence argument maps these kT-scale results onto larger-diameter cryogenic hydrogen jets.}


\begin{document}




\section{Introduction}
Intense, ultra-short laser pulses interacting with overdense matter produce some of the most extreme states of high-energy-density physics accessible in the laboratory: relativistic electron populations, TV/m electrostatic fields, and converging ion beams with MeV-scale energies, all generated on sub-picosecond timescales. Understanding the fundamental mechanisms that govern laser-plasma coupling, hot-electron transport, and laser-driven ion acceleration in this regime is of intrinsic scientific interest and directly relevant to compact laser-driven secondary sources such as pitcher-catcher or colliding-beam neutron generators, where efficient ion acceleration and localized convergence set the achievable yields \cite{NeutronSource1, NeutronSource2}.

Cryogenic hydrogen cylinders are a particularly well-controlled target geometry for isolating these mechanisms. Operating in a low-compression regime (in our simulations reaching on the order of $10\times$ solid density) keeps the target sufficiently transparent for high-repetition-rate systematic scans and femtosecond-resolution diagnostics, as already demonstrated at the European XFEL \cite{yang2025scalingwirecylindricalcompression} and at DRACO \cite{DRACO_Publ-Id:43136/1}, and as planned at the upcoming PEnELOPE facility \cite{Albach_Loeser_Siebold_Schramm_2019}. These conditions also expose physics that standard radiation-hydrodynamic (rad-hydro) codes fundamentally cannot access: the laser drives an intrinsically non-quasi-neutral electrostatic double-layer that sustains a TV/m radial field over hundreds of femtoseconds, producing a separate fast-ion population through direct reflection off this moving structure. Neither the extreme charge separation nor the resulting non-thermal particle channel can be recovered by fluid-averaged, single-temperature models. Capturing them requires a fully kinetic treatment.

In this work we present a predictive comparative study based on fully kinetic two-dimensional (2D3V) PIConGPU simulations \cite{PIConGPU2013}, establishing a numerical baseline for upcoming experimental campaigns at the Helmholtz-Zentrum Dresden-Rossendorf (HZDR). By directly contrasting the operational DRACO ($\tau = 30$~fs) and the upcoming PEnELOPE ($\tau = 150$~fs) drivers at two intensities ($a_0 = 12.7$ and $22.0$), we isolate the role of pulse duration in the particle kinematics, the structure of the charge-separation front, and the convergence of the driven compression wave. We show that both drivers produce a robust kinematic bifurcation into a fast (1--5~MeV) and a slow bulk (1--100~keV) ion population, and we derive a simple $2v_{hb}$ reflection model from the spatiotemporal electric-field data that quantitatively predicts the fast-ion energy across the full parameter scan.

We then explore the influence of an external axial magnetic field, scanning from a lab-achievable 20~T seed field up to 10~kT. By tracking the charge-separation field, the ion energy spectra, a compression-time diagnostic, and the peak core density across this scan, we identify the thresholds at which the hot-electron population becomes magnetised on the target scale, the laser-driven ion-acceleration mechanism is progressively quenched, and the net-inward compression phase is extended by up to a factor of two. We argue that the kT-scale fields used here are kinematically equivalent---in the ratio of hot-electron Larmor radius to target radius---to much more modest applied fields on the larger-diameter cryogenic hydrogen jets that could be produced experimentally, so that the reported physics is not confined to the specific 15~$\mu$m geometry simulated.

The central claim of this work is twofold: that the front-surface ion-acceleration channel in this regime is driven by a non-quasi-neutral electrostatic double layer whose instantaneous velocity quantitatively sets the fast-ion energies through a simple $2v_{hb}$ reflection law across both pulse durations and intensities studied; and that an applied axial magnetic field, once it brings the MeV hot-electron Larmor radius below the target radius, coherently quenches this acceleration channel, extends the net-inward compression phase, and---through a Larmor-radius-to-target-radius equivalence---becomes accessible at modest field strengths on larger-diameter targets.

\begin{figure}[htbp]
\begin{adjustwidth}{-\extralength}{0cm}
  \centering

  {\small\textbf{$\tau = 30$~fs}}\\[2pt]
  \begin{tabular}{@{}cccc@{}}
    \includegraphics[width=0.24\linewidth]{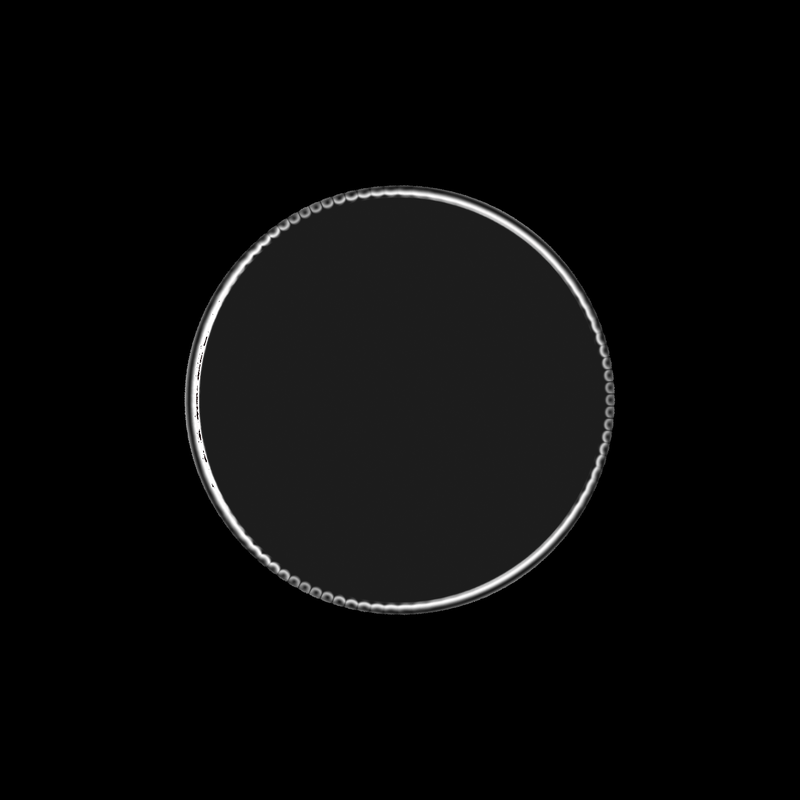} &
    \includegraphics[width=0.24\linewidth]{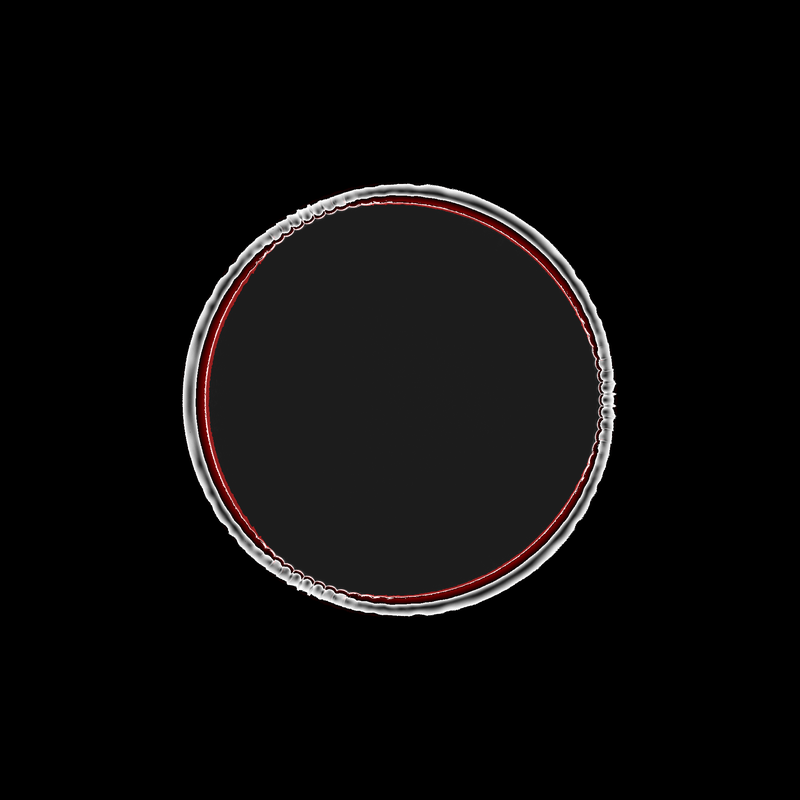} &
    \includegraphics[width=0.24\linewidth]{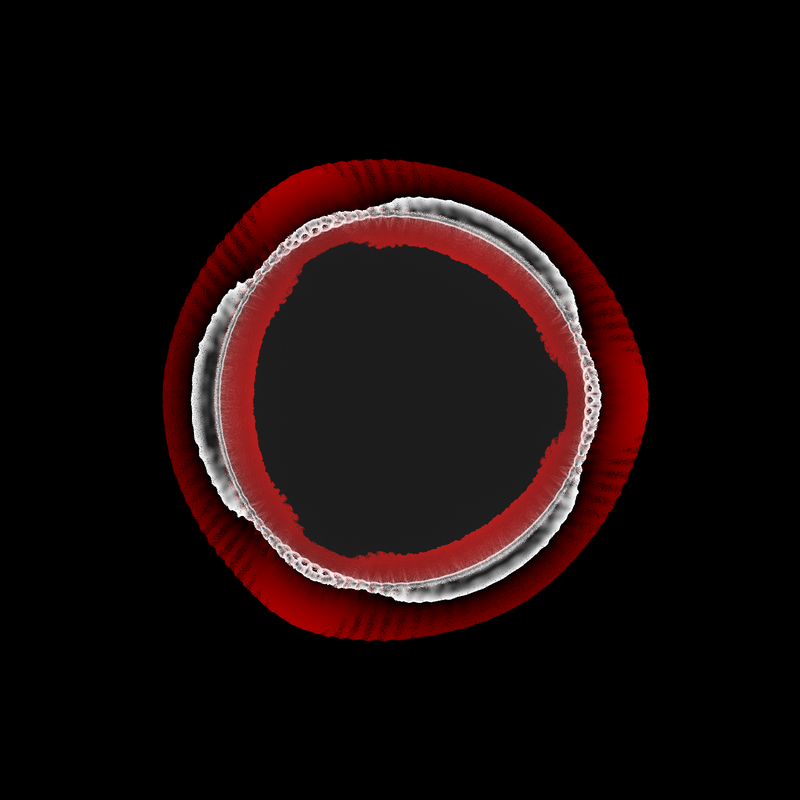} &
    \includegraphics[width=0.24\linewidth]{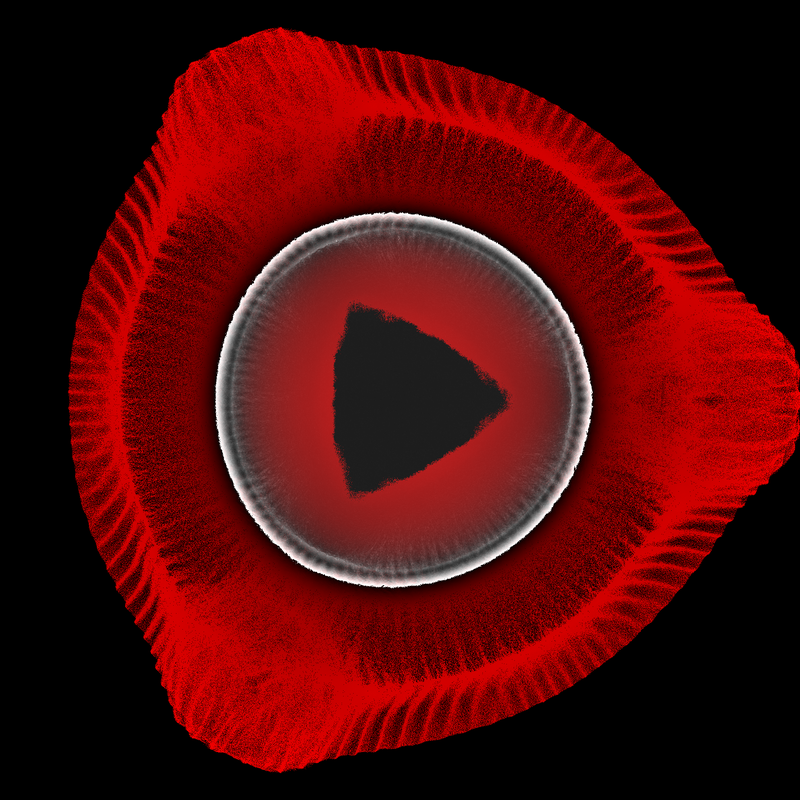} \\
    {\footnotesize $t = -30$~fs} & {\footnotesize $t = 0$~fs} & {\footnotesize $t = 100$~fs} & {\footnotesize $t = 250$~fs} \\[4pt]
    \includegraphics[width=0.24\linewidth]{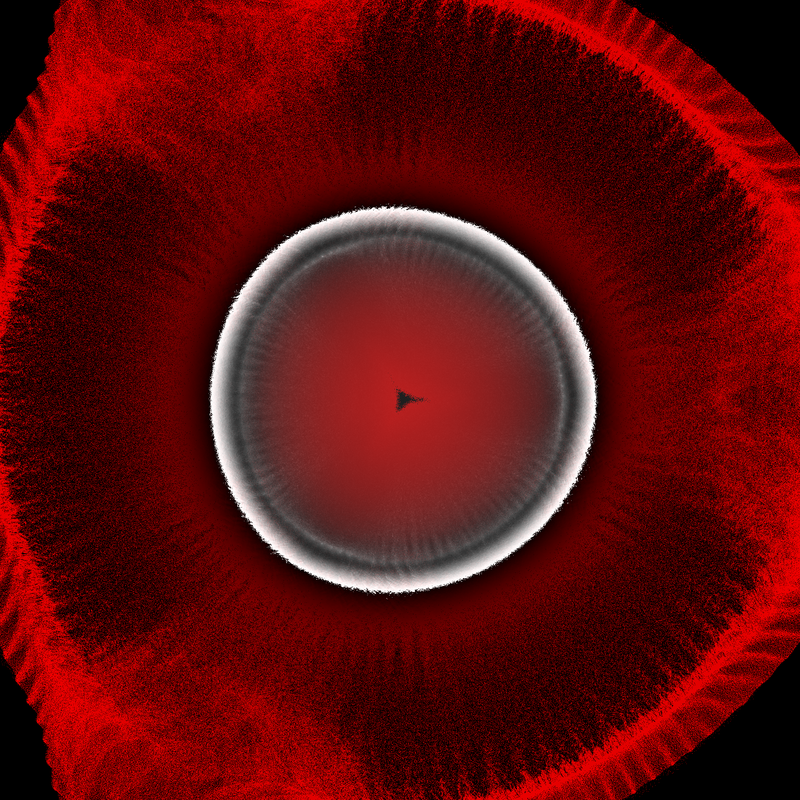} &
    \includegraphics[width=0.24\linewidth]{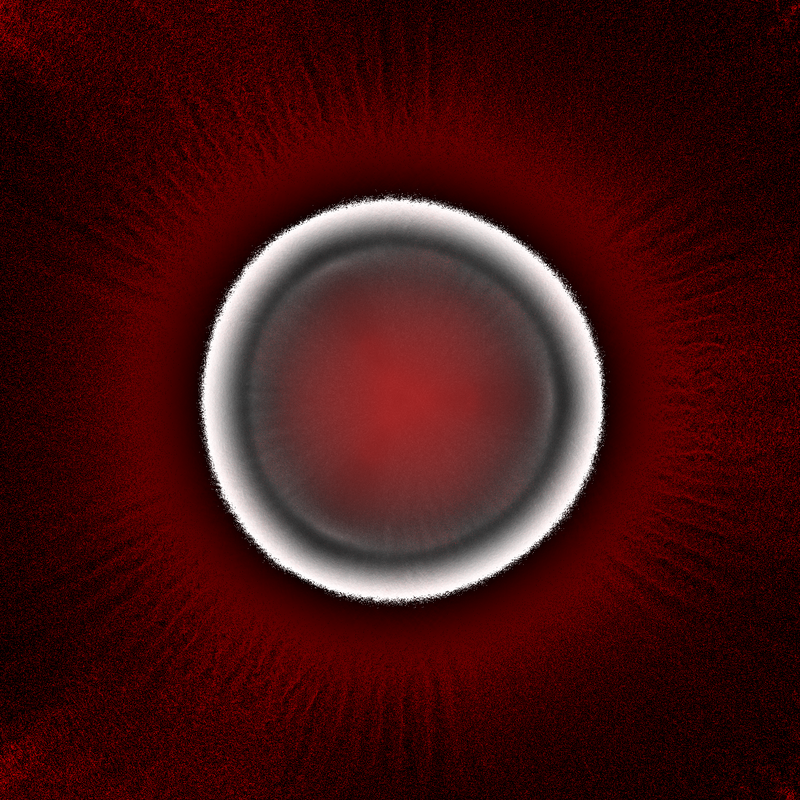} &
    \includegraphics[width=0.24\linewidth]{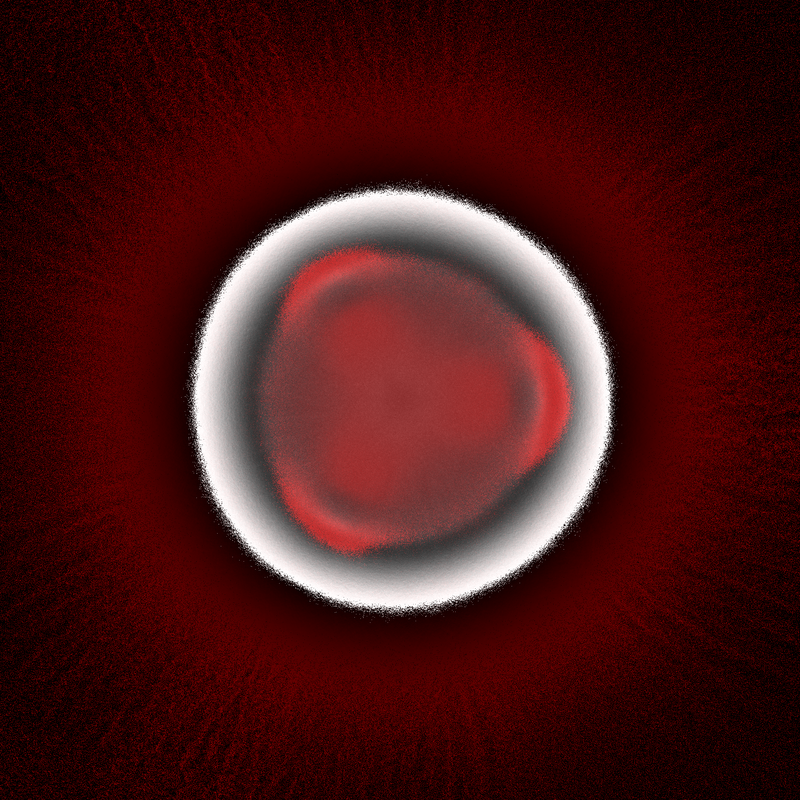} &
    \includegraphics[width=0.24\linewidth]{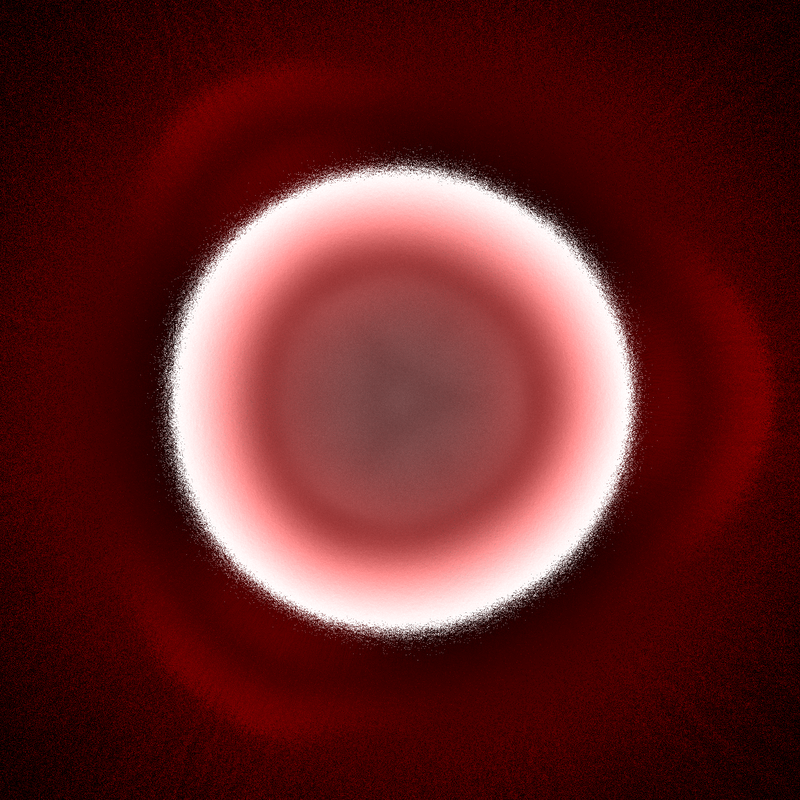} \\
    {\footnotesize $t = 400$~fs} & {\footnotesize $t = 500$~fs} & {\footnotesize $t = 650$~fs} & {\footnotesize $t = 1000$~fs} \\
  \end{tabular}

  \vspace{6pt}

  {\small\textbf{$\tau = 150$~fs}}\\[2pt]
  \begin{tabular}{@{}cccc@{}}
    \includegraphics[width=0.24\linewidth]{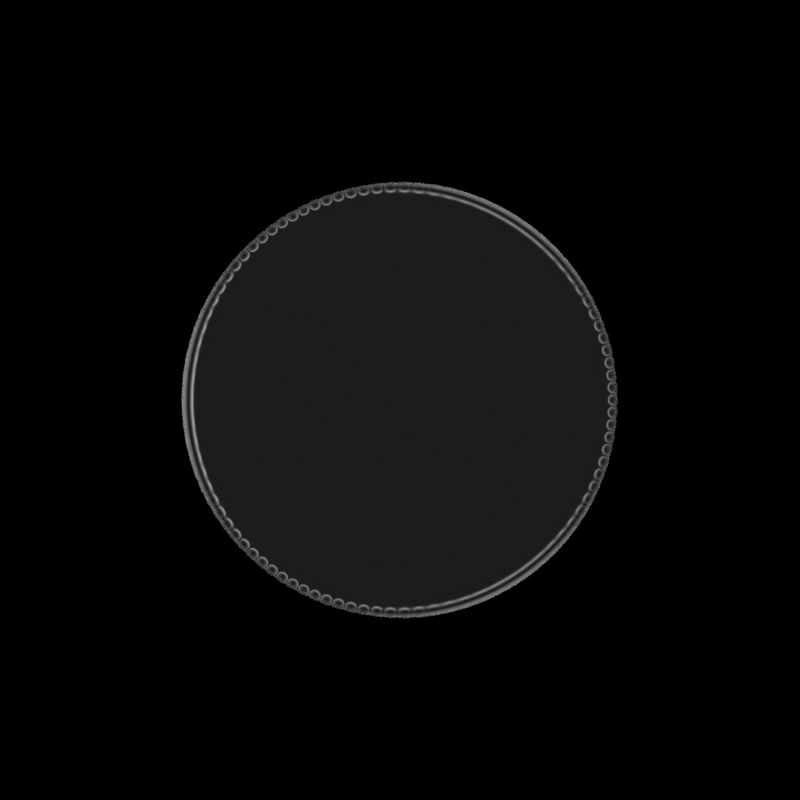} &
    \includegraphics[width=0.24\linewidth]{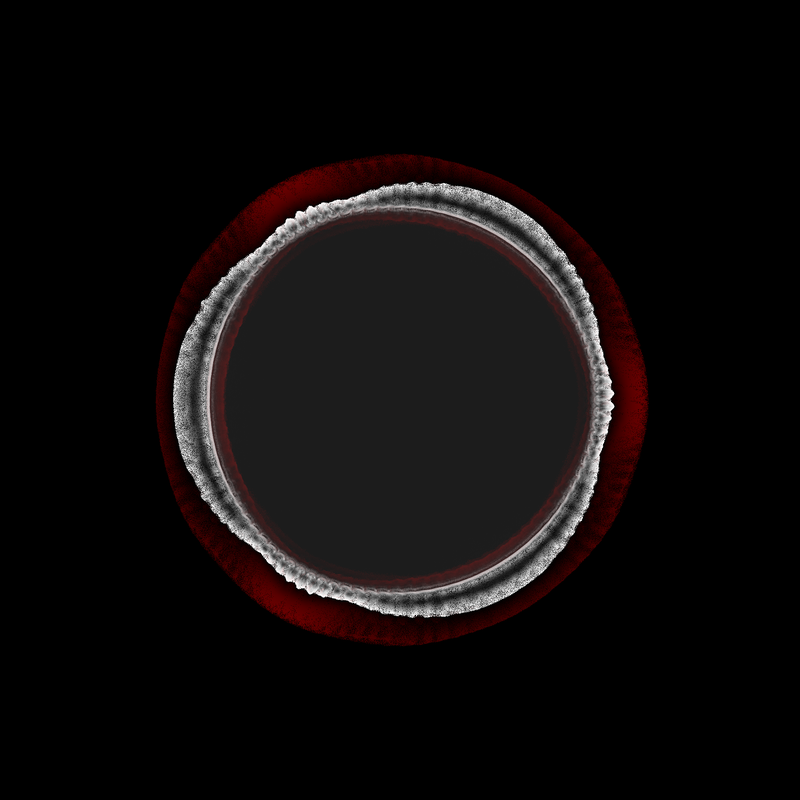} &
    \includegraphics[width=0.24\linewidth]{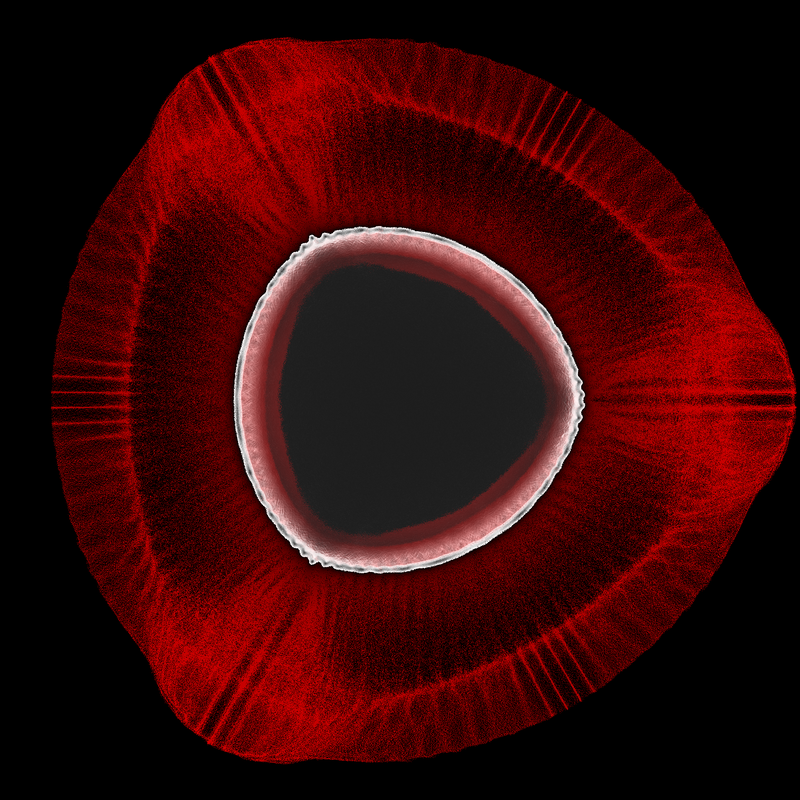} &
    \includegraphics[width=0.24\linewidth]{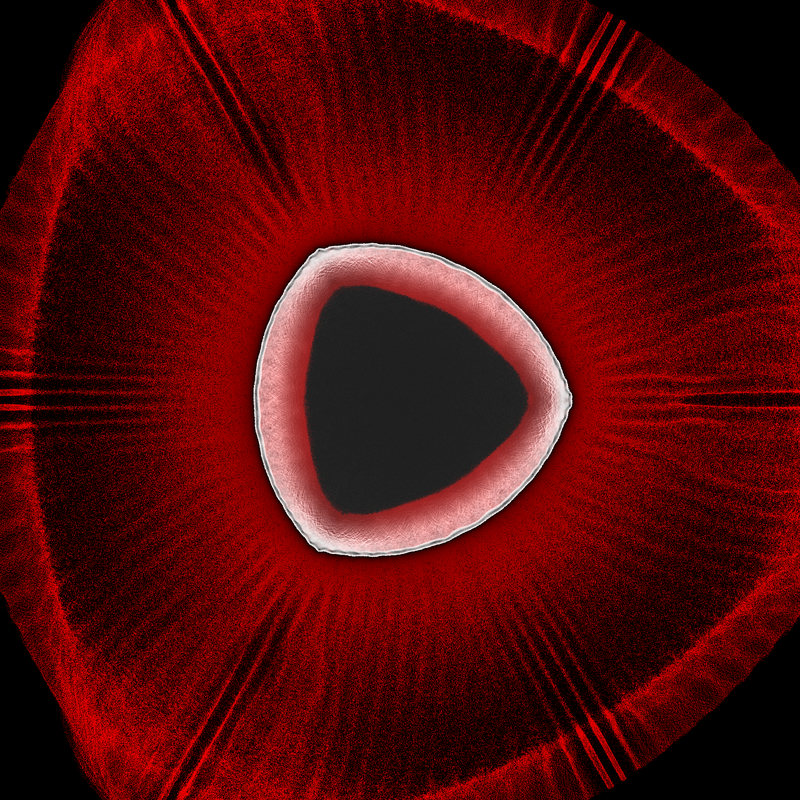} \\
    {\footnotesize $t = -500$~fs} & {\footnotesize $t = -250$~fs} & {\footnotesize $t = 0$~fs} & {\footnotesize $t = 100$~fs} \\[4pt]
    \includegraphics[width=0.24\linewidth]{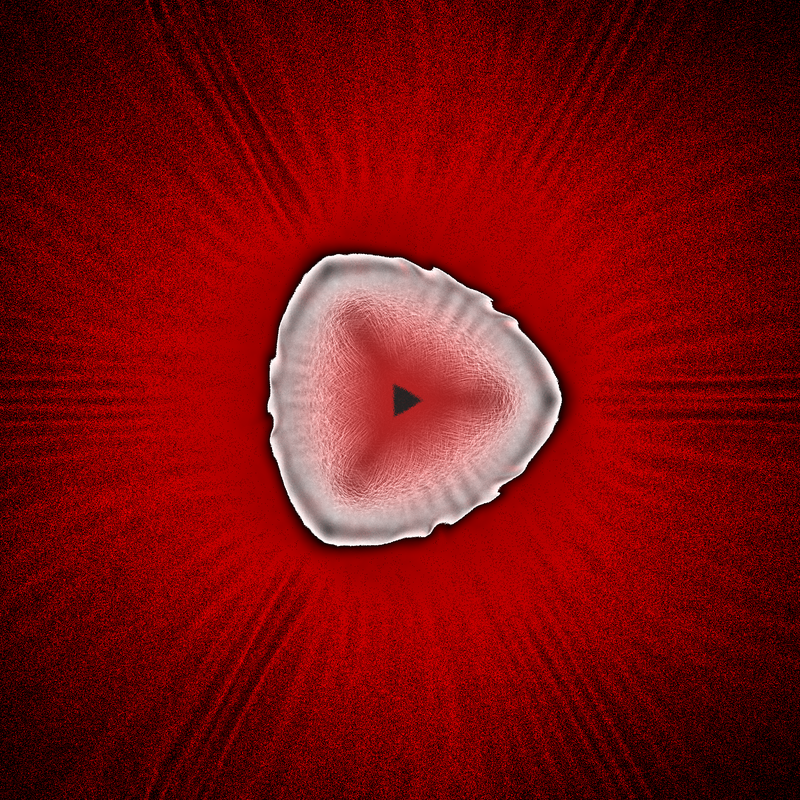} &
    \includegraphics[width=0.24\linewidth]{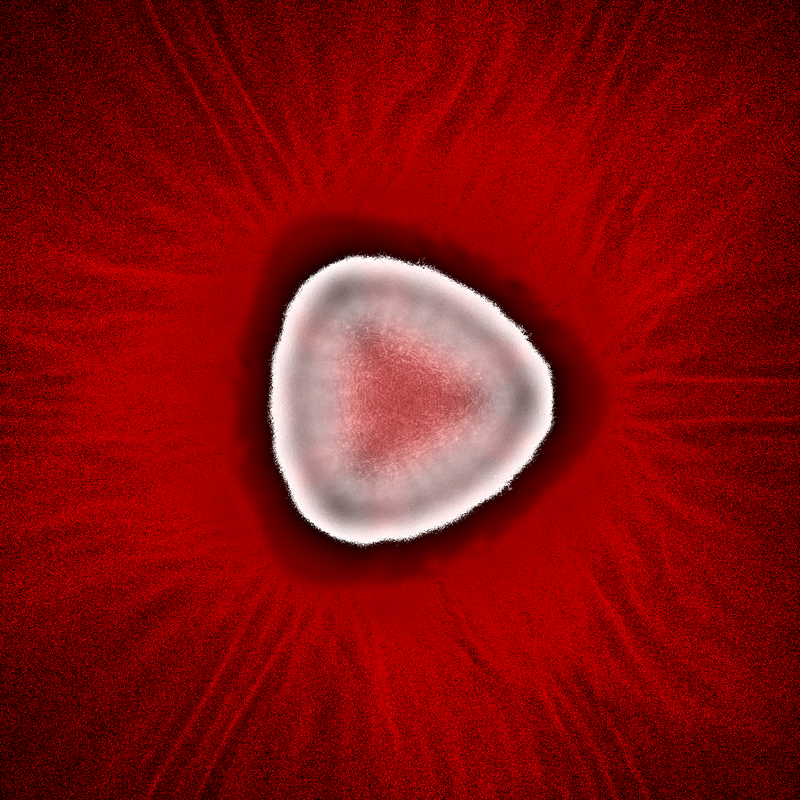} &
    \includegraphics[width=0.24\linewidth]{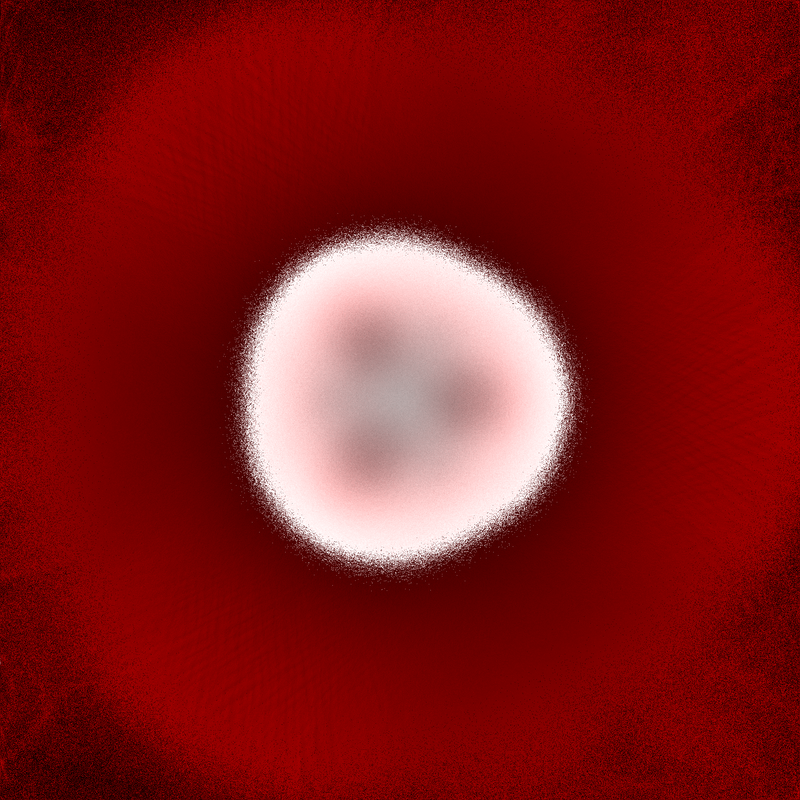} &
    \includegraphics[width=0.24\linewidth]{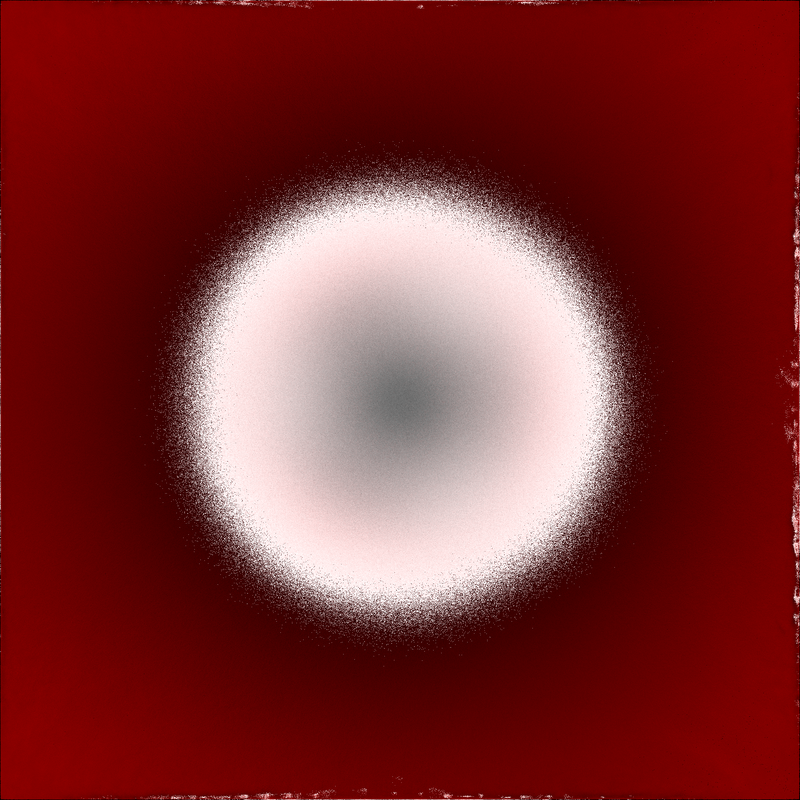} \\
    {\footnotesize $t = 300$~fs} & {\footnotesize $t = 450$~fs} & {\footnotesize $t = 1000$~fs} & {\footnotesize $t = 1500$~fs} \\
  \end{tabular}

  \caption{Evolution of the target kinetic energy density for the $\tau = 30$~fs (top block) and $\tau = 150$~fs (bottom block) drivers, each shown as eight snapshots arranged in a $2 \times 4$ time sequence (reading left-to-right, top-to-bottom). Frame times are quoted in femtoseconds relative to the arrival of the peak laser intensity at the target surface ($t = 0$~fs); negative times denote the rising edge before that peak. The colour scale is deliberately discontinuous: it encodes the cold/warm bulk and the hot tail on a single frame. Particles with kinetic energy below $100$~keV are shown on a monochrome black-to-white ramp (black at $0$~keV, white at $100$~keV), so the bright-white regions mark the $\sim 100$~keV bulk ion population. At $100$~keV the scale resets to black and a second, red-hued ramp (black at $100$~keV $\rightarrow$ bright red at $10$~MeV) takes over, so that any red pixel unambiguously flags a population in the MeV-class fast-ion band identified in Figures~\ref{fig:histograms} and~\ref{fig:kinematics22}. The sequences illustrate the symmetric piston effect at the target surface, the launching of converging radial compression waves, and the eventual core disintegration.}
  \label{fig:kinetic_map}
  \end{adjustwidth}
\end{figure}

\begin{adjustwidth}{-\extralength}{0cm}
\section{Simulation Setup}

To investigate the interaction dynamics of ultra-short laser pulses with solid hydrogen, we performed a series of two-dimensional (2D3V) kinetic simulations using the fully relativistic, electromagnetic, open-source particle-in-cell (PIC) code PIConGPU \cite{PIConGPU2013}. All simulations were performed on the JUPITER Booster \cite{JUPITER2025} system.

This fully kinetic approach natively captures physical mechanisms that are typically inaccessible or require heuristic approximations in standard radiation-hydrodynamic modeling. First, the explicit Maxwell solver inherently accounts for coherent beam superposition. Consequently, the three-beam direct-drive configuration (Section~\ref{sec:laser_configuration}) captures beam-beam interference, polarization-dependent coupling, wave-based refraction, and transient relativistic transparency at the time-evolving critical surface.

Second, advancing both electron and ion species via the Vlasov-Maxwell system avoids the need for fluid heat-flux closures. Coupled with the FLYonPIC collisional-radiative atomic-physics module \cite{Marre2026}, this allows for the self-consistent modeling of non-LTE atomic kinetics and highly non-local electron transport. These kinetic capabilities are critical for the magnetized regime detailed in Section~\ref{sec:magnetic_results}, where transport is anisotropic on the electron mean-free-path scale and the cyclotron-to-laser frequency ratio approaches and surpasses unity. 

Finally, the simulated time window of $2.6$~ps extends beyond the sub-picosecond electron-equilibration phase well into the ion-response regime. Because fully kinetic studies reaching this multi-picosecond temporal scale remain rare, these results uniquely complement the existing rad-hydro literature.

\subsection{Numerical Domain and Target Properties}\label{sec:numerical_domain}
The simulation domain spans $30.3 \times 30.3$~$\mu$m$^2$ ($10112 \times 10112$ cells) and is resolved with a uniform Cartesian grid. The spatial resolution is set to $\Delta x = \Delta y = 3$~nm, which sufficiently resolves the plasma skin depth at the densities under consideration. The temporal resolution is determined by the Courant-Friedrichs-Lewy condition, yielding a timestep of $\Delta t \approx 5.5$~as. Particle dynamics are solved using the Higuera-Cary particle pusher\cite{Higuera2017}, while the current deposition is handled via the charge-conserving Esirkepov scheme\cite{ESIRKEPOV2001144}. At the center of the domain, we initialize a solid-density cryogenic hydrogen cylinder (based on \cite{cryoH2}). The target has a radius of $R = 7.5$~$\mu$m and an initial neutral density of $n_0 = 5.24 \times 10^{28}$~m$^{-3}$. To accurately capture the laser-plasma coupling at the leading edges of the pulses, field and collisional ionization processes are dynamically tracked throughout the interaction using PIConGPU's atomic physics extension FLYonPIC \cite{Marre2026}. Throughout this work, any quantity reported in per-volume units (e.g.\ particle densities in cm$^{-3}$ or kinetic energy densities in J\,cm$^{-3}$) is obtained from the native 2D per-area output of the simulation by assuming an implicit cell depth equal to the in-plane resolution, $\Delta z = \Delta x = 3$~nm; equivalently, the 2D3V run is interpreted as a thin slab of thickness $\Delta z$ in the out-of-plane direction. The 3D initial neutral density $n_0 = 5.24 \times 10^{28}$~m$^{-3}$ quoted above is the underlying physical target density used to initialize the macroparticle distribution and is consistent with this convention.
\subsection{Laser Configuration}\label{sec:laser_configuration}
To mimic a direct-drive geometry, the target is symmetrically irradiated by three converging laser pulses. The beams are offset by $120^\circ$ and propagate inward toward the central axis. Each pulse is linearly polarized with a central wavelength of $\lambda = 800$~nm and features a Gaussian transverse profile with a waist of $w_0 = 7.5$~$\mu$m, matched directly to the target radius.

To systematically compare the operational regimes of the Draco and Penelope laser systems, the pulses were modeled with Gaussian temporal profiles defined by standard deviation parameters of $\tau = 30$~fs and $150$~fs. These yield true full-width at half-maximum (FWHM) intensity durations broader by $2\sqrt{\ln 2} \approx 1.665$. The $\tau = 30$~fs pulse serves as an archetype for the impulsive, ultra-short Draco-class driver, while the $\tau = 150$~fs pulse represents the sustained Penelope-class driver. For consistency with the numerical setup, these configurations are referred to throughout the text and figures by their base $\tau$ values (30~fs and 150~fs). For each temporal configuration, we simulate peak normalized vector potentials of $a_0 = 12.7$ and $a_0 = 22.0$. These parameters correspond to peak intensities on the order of $3.45\times10^{20}$ to $10^{21}$~W/cm$^2$, placing the interaction firmly within the highly relativistic regime for electron transport. The total simulation time is 2.6~ps, allowing sufficient time to observe both the laser-plasma interaction and the subsequent convergence of the driven compression waves. Unless explicitly stated otherwise, the time axis of every diagnostic plot in this work is referenced to the arrival of the peak laser intensity at the target surface, so that $t = 0$~ps marks the instant of peak on-target intensity and negative times denote the rising edge of the pulse before that peak.

We calculate the laboratory-equivalent pulse energy---the total energy of a 3D Gaussian beam in vacuum sharing the simulation's peak intensity $I_0$, transverse waist $w_0$, and duration $\tau$. Peak intensity relates to the normalized amplitude via $I_0 \approx 1.37 \times 10^{18}\,(a_0/\lambda_{\mu\mathrm{m}})^2$~W\,cm$^{-2}$. Assuming a rotationally symmetric spatial profile $I(r) = I_0 \exp(-2 r^2 / w_0^2)$ and a Gaussian temporal envelope $I(t) = I_0 \exp(-t^2/\tau^2)$ (where $\tau_{\mathrm{FWHM}} \approx 1.665\,\tau$), the separable space-time integral yields a single-beam energy of:
\begin{equation}
E_{\mathrm{beam}} \;=\; I_0\,\frac{\pi w_0^2}{2}\,\tau\sqrt{\pi}.
\label{eq:laser_pulse_energy}
\end{equation}
When $I_0$ is expressed in W\,cm$^{-2}$, $w_0$ in cm, and $\tau$ in seconds, Equation~\eqref{eq:laser_pulse_energy} gives the energy in joules. For our three-beam setup, the total laboratory-equivalent energy is $E_{\mathrm{tot}} = 3 E_{\mathrm{beam}}$. Note that this physical volume integration is entirely independent of the diagnostic slab thickness $\Delta z$ (Section~\ref{sec:numerical_domain}) used by the 2D code to evaluate volumetric densities. The corresponding laboratory-equivalent energies for the driving configurations used in this work ($w_0 = 7.5~\mu$m, $\lambda = 800$~nm) are summarized in Table~\ref{tab:laser_energy}.
\end{adjustwidth}
\begin{table}[hbt!]
\begin{adjustwidth}{-\extralength}{0cm}
\centering
\begin{tabular}{cc|cc|cc}
\hline\hline
$a_0$ & $I_0$ (W\,cm$^{-2}$) & $\tau$ (fs) & $\tau_{\mathrm{FWHM}}$ (fs) & $E_{\mathrm{beam}}$ (J) & $E_{\mathrm{tot}}$ (J) \\ \hline
12.7 & $3.45 \times 10^{20}$ & 30  & $\sim 50$  & 16  & 49  \\
12.7 & $3.45 \times 10^{20}$ & 150 & $\sim 250$ & 81  & 240 \\
22.0 & $1.04 \times 10^{21}$ & 30  & $\sim 50$  & 49  & 150 \\
22.0 & $1.04 \times 10^{21}$ & 150 & $\sim 250$ & 240 & 720 \\
\hline\hline
\end{tabular}
\caption{Laboratory-equivalent laser energies for the parameters used in the 2D simulations, assuming a 3D Gaussian focal spot with $w_0 = 7.5~\mu$m and $\lambda = 800$~nm.}
\label{tab:laser_energy}
\end{adjustwidth}
\end{table}
\begin{adjustwidth}{-\extralength}{0cm}

\subsection{External Magnetic Field}
To evaluate the threshold for magnetic confinement of hot electrons and its impact on core energy deposition, an external, static axial magnetic field ($B_z$) is applied perpendicular to the 2D simulation plane. We use a single-particle, target-scale notion of \emph{magnetisation}: a species is \emph{magnetised on the target scale} at kinetic energy $E_k$ when its relativistic Larmor radius $r_L$ in the applied field satisfies $r_L < R$, where $R = 7.5~\mu$m is the initial cylinder radius (Section~\ref{sec:numerical_domain}). Consequently the typical particle completes a substantial fraction of a cyclotron orbit before crossing the target and transverse motion is dominated by gyration rather than by ballistic streaming across $R$. Numerical values of $r_L(E_k,B_z)$ follow Equation~\eqref{eq:larmor} and are tabulated in Table~\ref{tab:larmor}.

For each laser configuration, the simulation is repeated across four magnetic field strengths: 0~T (baseline), 20~T (representing a lab-achievable seed field), and 1~kT to 10~kT. While the kT-scale fields are experimentally unachievable for this specific 15~$\mu$m geometry, they serve as a proxy for the magnetised physics of larger targets at more modest applied fields: since the relevant quantity is the ratio of the particle Larmor radius to the target radius, holding $r_L/R$ constant requires $B \propto 1/R$, so 10~kT on a 7.5~$\mu$m target is kinematically equivalent---in terms of hot-electron magnetisation on the target scale---to $\sim$10~T on a 7.5~mm-diameter cryogenic hydrogen jet. This mapping is a geometric kinematic equivalence; it does not constitute exact plasma similarity (which would additionally require density to scale as $1/R^2$), but it captures the onset of the confinement and transport-suppression effects that are the subject of Section~\ref{sec:magnetic_results}.

At our solid density, the mean inter-particle spacing is $d \approx n_0^{-1/3} \approx 0.27$~nm, so even in the strongest-field cases the electron gyroradius values in Table~\ref{tab:larmor} exceed $d$ by several orders of magnitude. At the same time, the electron cyclotron frequency reaches $\omega_{ce} \approx 1.76\times10^{15}$~rad/s at 10~kT, which is larger than typical electron-ion collision frequencies for keV-class bulk electrons at solid density ($\nu_{ei} \sim 10^{13}$~s$^{-1}$), placing the target in a strongly magnetised transport regime.

\section{Results Part I: Ion Acceleration in the Unmagnetized Regime}
To establish a baseline for evaluating the influence of external magnetization and to compare the kinematics of different laser drivers, we first analyze the unmagnetized ($B_z = 0$) case. In this regime, the target dynamics are governed entirely by the localized laser-plasma interactions at the critical density surface and the subsequent evolution of the driven compression waves.
\subsection{Macroscopic Target Dynamics and Shock Convergence}
The symmetric irradiation of the 15~$\mu$m solid-density cryogenic hydrogen cylinder by three converging laser pulses initiates rapid plasma formation and expansion at the target surface. Figure~\ref{fig:kinetic_map} illustrates the macroscopic evolution of the target's kinetic energy density over time. 
\end{adjustwidth}
\begin{adjustwidth}{-\extralength}{0cm}
Upon incidence, the intense ponderomotive pressure of the laser initiates a hole-boring phase. The localized energy deposition launches distinct radial compression waves that propagate inward toward the central axis. While both the 30~fs (DRACO-class) and 150~fs (PEnELOPE-class) pulses successfully drive converging compression fronts, the temporal duration of the driver fundamentally alters the underlying particle kinematics, necessitating a detailed examination of the localized electric fields. Due to cylindrical symmetry, the subsequent quantitative analyses are azimuthally averaged over the polar angle. This isolates the purely radial dependence of the interaction dynamics.
\subsection{Spatiotemporal Evolution of the Charge-Separation Front}
The physical driver of the inward ion acceleration is the intense electrostatic field generated by laser-driven charge separation. Driven by the ponderomotive force ($F_p \propto -\nabla I$) the laser acts as a macroscopic piston, displacing electrons inward away from the highest intensity regions to establish a moving electrostatic double-layer. This double-layer functions as a moving capacitor, consisting of a leading compressed electron sheet and a trailing positive ion cavity; consequently, the macroscopic electric field is entirely contained within this gap, neutralized on the outside by both the parallel-plate geometry and the plasma's innate Debye shielding.
\end{adjustwidth}
\begin{figure}[htbp] 
\begin{adjustwidth}{-\extralength}{0cm}
    \centering
    \includegraphics[width=1\linewidth]{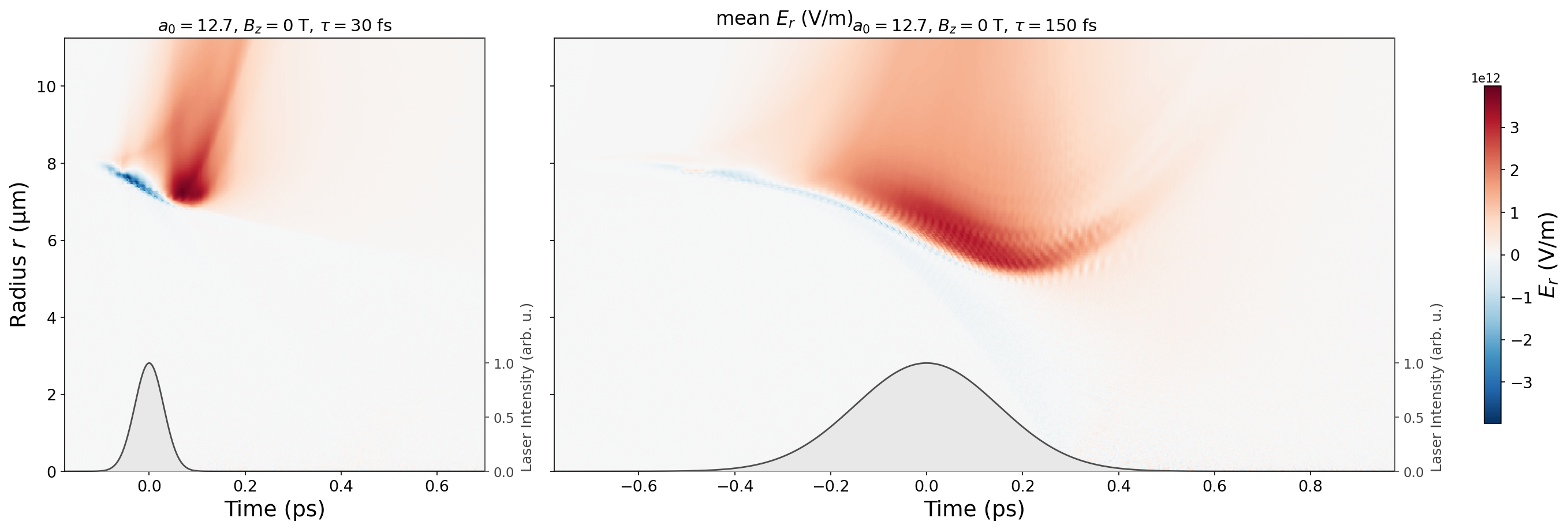}
    \caption{Evolution of the radial electric field ($E_r$) for 30~fs (left) and 150~fs (right) pulses at $a_0 = 12.7$. The blue regions indicate inward-pointing fields, while red regions indicate outward-pointing fields. The shaded grey curves at the bottom depict the temporal laser intensity envelope, where $t = 0$~ps corresponds to the arrival of the peak intensity at the target surface.}
  \label{fig:electric_fields}
  \end{adjustwidth}
\end{figure}
\begin{adjustwidth}{-\extralength}{0cm}
Figure~\ref{fig:electric_fields} captures this moving charge-separation front, revealing radial electric fields ($E_r$) on the order of several TV/m ($\sim 3 \times 10^{12}$~V/m). Because the ponderomotive force relies on the spatial gradient of the laser intensity, macroscopic charge separation is strictly driven along the rising edge of the pulse. This gradient dependence fundamentally dictates the stark contrast observed between the two laser drivers. The ultra-short 30~fs pulse compresses its intensity rise into a much steeper spatial envelope than the 150~fs pulse. This steeper gradient exerts a significantly stronger ponderomotive force, which in turn generates a higher maximum electric field. However, due to its brevity, the charge-separation front under 30~fs irradiation is highly transient; the intense inward-pointing field (blue) flashes and dissipates rapidly, penetrating only shallowly into the target and characterizing the interaction as a sharp, impulsive kick.

Conversely, the 150~fs pulse sustains the ponderomotive pressure for a significantly longer duration. The corresponding $E_r$ plot demonstrates a robust charge-separation front that maintains its TV/m amplitude for hundreds of femtoseconds, actively boring deeper into the target radius. As this electrostatic double-layer is driven inward, it continuously performs work on the local ion population.

\subsection{Ion Energy Spectra and Kinematic Bifurcation}\label{sec:energy_Histograms_introduction}
The differing lifespans of these charge-separation fronts directly dictate the final ion energy spectra \cite{Robinson_2009}. The inward propagation of the intense electric field acts as a moving potential ramp, creating a distinct kinematic bifurcation in the ion population: it separates the plasma into a fast, ballistic stream and a slower, entrained bulk.
\end{adjustwidth}
\begin{figure}[htbp]
    \begin{adjustwidth}{-\extralength}{0cm}
      \centering
      \includegraphics[width=1\linewidth]{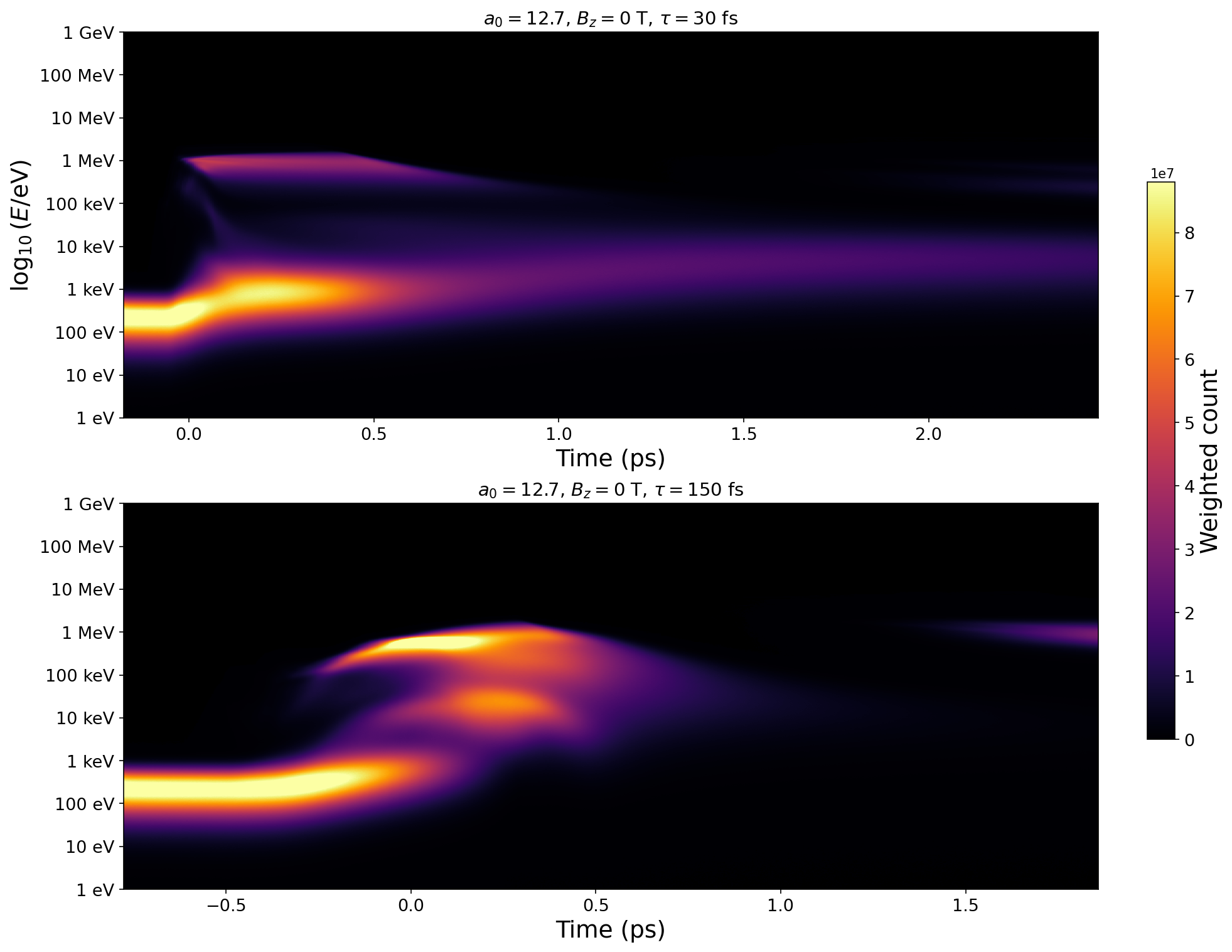}
      \caption{Energy histograms of ions with negative (inward) radial momentum for the 30~fs (top) and 150~fs (bottom) pulses at $a_0 = 12.7$. Under 30~fs irradiation, only a small fraction of the ions are accelerated to high energies. In contrast, for the 150~fs pulse, fast-moving ($\sim 1$~MeV) ions constitute the majority of the inward-directed particles. The distributions highlight two distinct populations: the bulk ions ($10$--$100$~keV) heated and accelerated via radiation pressure, and the fast ions ($\sim 1$~MeV) reflected directly from the electrostatic double-layer. Readers can compare the full, omnidirectional energy histograms via the interactive Simulation Data Viewer \cite{Optolowicz2026}.}
      \label{fig:histograms}
  \end{adjustwidth}
\end{figure}
\begin{adjustwidth}{-\extralength}{0cm}
As shown in the inward radial momentum histograms (Figure~\ref{fig:histograms}), both laser configurations generate a fast ion population reaching $\sim 1$~MeV. These ions represent a fraction of the background fuel that is strongly accelerated by reflecting off the intense electrostatic field. However, their spectral signatures differ fundamentally due to the kinematics of this moving double-layer. Under the impulsive 30~fs irradiation, the ponderomotive push is exceedingly brief, causing the double-layer to advance inward at a roughly constant velocity during the interaction. Consequently, the ions reflected off this front acquire a uniform energy, producing the flat, constant-energy trace in the 1~MeV band (followed by a sharp time-of-flight cutoff as the ions cross the target center at $r=0$ and their radial momentum reverses). In contrast, the sustained 150~fs pulse applies a ponderomotive force over a much longer rising edge, extending from roughly $-0.5$~ps up to the laser peak at $0$~ps. During this extended period, the double-layer continuously accelerates to higher inward speeds, as seen in Figure~\ref{fig:electric_fields}. Because the reflection energy depends directly on the instantaneous speed of the advancing front, this acceleration yields a non-monoenergetic distribution, causing the fast-ion population to exhibit a pronounced energy sweep from hundreds of keV up to a few MeV.

The rest of the ions lack the relative potential to be fully reflected. These ions are entrained within the moving density wave, forming the slower hole-boring front. Within this bulk population ($1$--$100$~keV) the difference between the impulsive and sustained drivers is:

\begin{itemize}
\item \textbf{Impulsive Acceleration and Ablation Dynamics (30 fs):} Once the 30~fs pulse ends, the confining ponderomotive pressure vanishes. The macroscopic target evolution is subsequently dominated by ablation dynamics. Without the sustained inward laser push, the macroscopic ablation and decompression of the plasma smoothly halt the inward momentum of the bulk, causing the slow ions to reverse direction and fade from the inward-momentum diagnostic.

\item \textbf{Sustained Piston Dynamics (150 fs):} The rising edge of the pulse pre-heats and accelerates the target bulk to $\sim 1$~keV. Upon reaching peak intensity, the sustained charge-separation front acts as an active piston, fighting off decompression and continuously driving the entrained bulk ions inward. This is evidenced by the distinct spectral jump in the histogram, where the bulk ion energy actively rises from 1~keV to around 80~keV over time. 

\end{itemize}

Figure~\ref{fig:kinematics12p7} shows both drivers simultaneously: the electric-field evolution and the resulting ion energy histograms with the $2v_{hb}$ prediction overlaid.

\end{adjustwidth}
\begin{figure}[htbp]
\begin{adjustwidth}{-\extralength}{0cm}
\centering
\includegraphics[width=1\linewidth]{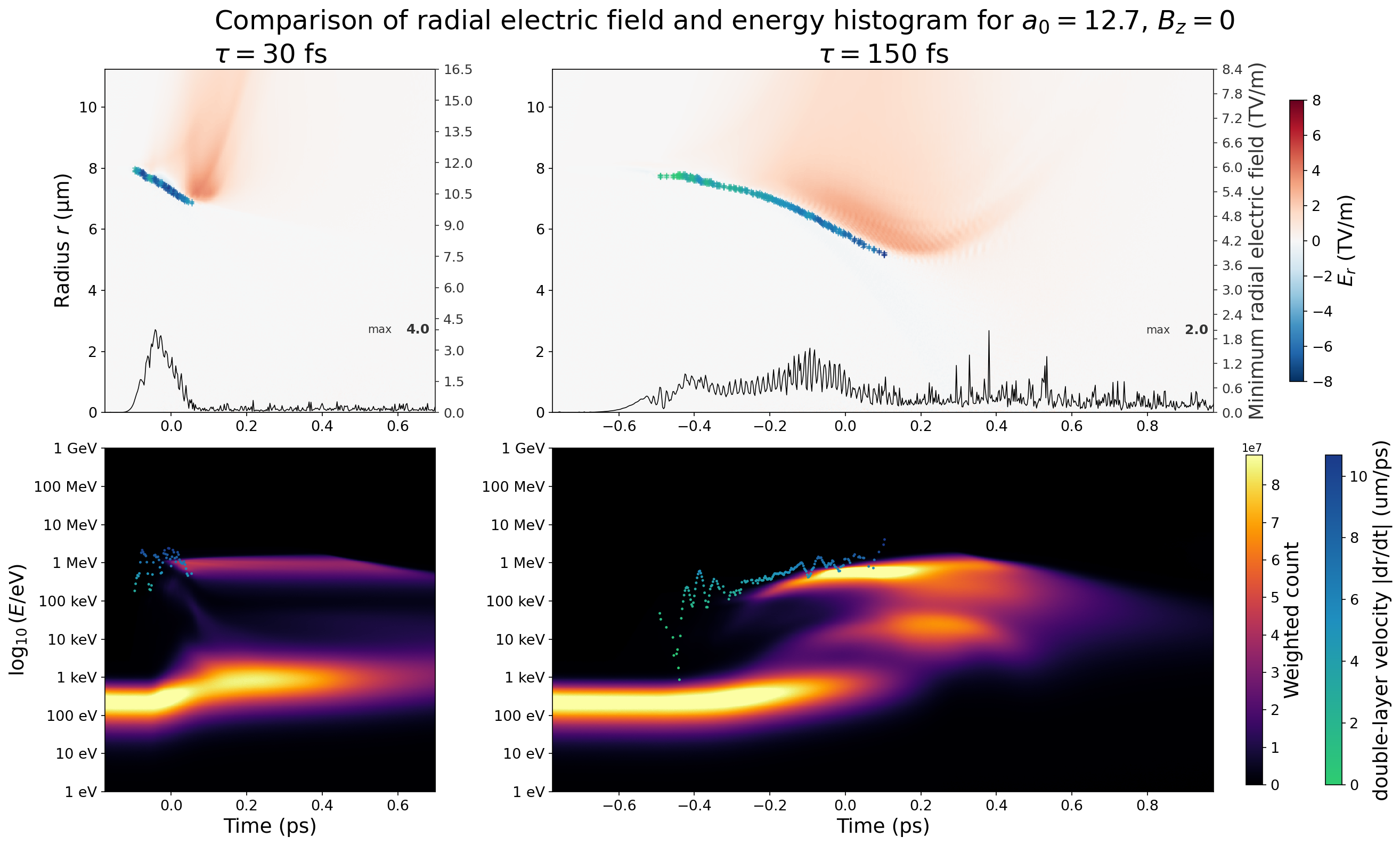}
\caption{Kinematic analysis of the 30~fs (left column) and 150~fs (right column) pulses at $a_0 = 12.7$, $B_z = 0$. (Top) $E_r(r,t)$ colourmap with overlaid dot markers tracking the radial trajectory of the charge-separation front (gated minimum $E_r$); the black curve below the colourmap shows the temporal evolution of the maximum accelerating field ($-\min E_r$), with its absolute peak annotated on the right-hand scale. (Bottom) Inward-ion energy histograms with the $2v_{hb}$ reflection-energy markers overlaid (calculation procedure outlined in the main text). Under the impulsive 30~fs driver the front moves at nearly constant speed, producing a flat $\sim 1$~MeV trace; under the sustained 150~fs driver the front accelerates continuously, sweeping the reflected-ion energy from hundreds of keV to $\sim 1$~MeV.}
\label{fig:kinematics12p7}
\end{adjustwidth}
\end{figure}
\begin{adjustwidth}{-\extralength}{0cm}

To mathematically link the observed charge-separation fields (Figure~\ref{fig:electric_fields}) directly to the resulting fast-ion spectra (Figure~\ref{fig:histograms}), we extract the propagation velocity of the charge-separation layer ($v_{hb}$) from the $E_r(r,t)$ data. At each timestep, we record the radial position of the minimum electric field, provided it exceeds a specific magnitude threshold. After applying spatial smoothing to this trajectory, the instantaneous velocity is calculated using a central finite-difference approximation and further refined with a 7-point temporal rolling mean (averaging 3 timesteps forward and backward) to suppress high-frequency numerical noise. Assuming an ideal reflection at the moving potential barrier, background protons gain a velocity of $2v_{hb}$ in the laboratory frame \cite{Robinson_2009}. Converting this reflected velocity into proton kinetic energy using the relativistic relation $E_k = (\gamma - 1)m_p c^2$ \cite{NRL_formulary_2019} yields the theoretical traces plotted in Figure~\ref{fig:kinematics12p7}. We adopt the same overlay convention on every inward-ion histogram in this work: each marker is placed at the instantaneous predicted reflection energy $E_k(2v_{hb}(t))$, with its colour encoding the velocity of the charge-separation front $v_{hb}(t)$, so the trace and its colour bar are two views of the same extracted velocity series. The overlaid prediction oscillates around the constant-energy fast-ion band produced by the impulsive 30~fs driver, and tracks the upper edge of the swept fast-ion band produced by the 150~fs driver, confirming that this population originates from reflection off the charge-separation front.

It is crucial to distinguish this charge-separation reflection mechanism from Target Normal Sheath Acceleration (TNSA) \cite{Snavely_2000, Wilks_2001}, which is frequently the dominant process for generating MeV-class ions in high-intensity interactions. Because our uncompressed cryogenic target is thick ($\ge 15$~$\mu$m in diameter), the strong rear-surface electrostatic sheath characteristic of TNSA cannot be effectively established. Instead, the front-surface ponderomotive hole-boring mechanism operates as the primary acceleration pathway. 

Scaling the incident intensity to $a_0 = 22.0$ ($I \approx 1.04 \times 10^{21} \text{W/cm}^2$) preserves the fundamental kinematic bifurcation of the ion populations but amplifies the localized electric fields and the resulting kinetic energy ranges. As detailed in Figure~\ref{fig:kinematics22}, the fast-ion band for both the 30~fs and 150~fs configurations shifts from $\sim 1$~MeV into the 1--5~MeV range; the $2v_{hb}$ prediction now tracks upper edge of the fast-ion band more precisely preserving the same overlay behaviour seen at $a_0 = 12.7$. Unlike at $a_0 = 12.7$ (Figure~\ref{fig:kinematics12p7}), the individual dot markers tracking the front's radial trajectory are omitted from the $E_r$ colourmap at this intensity to avoid obscuring the field structure. The calculated theoretical reflection energy based on this field trajectory is plotted directly onto the ion energy histograms. The elevated ponderomotive drive makes transient kinetic features more pronounced. For instance, the impulsive 30~fs pulse at $a_0 = 22.0$ triggers a distinct, immediate jump in the bulk ion energy from 1~keV to 10~keV, which then gradually declines to a few keV as the target undergoes ablation. The exact mechanisms governing this specific low-energy transition remain a subject for future investigation.

\end{adjustwidth}
\begin{figure}[htbp]
\begin{adjustwidth}{-\extralength}{0cm}
\centering
\includegraphics[width=1\linewidth]{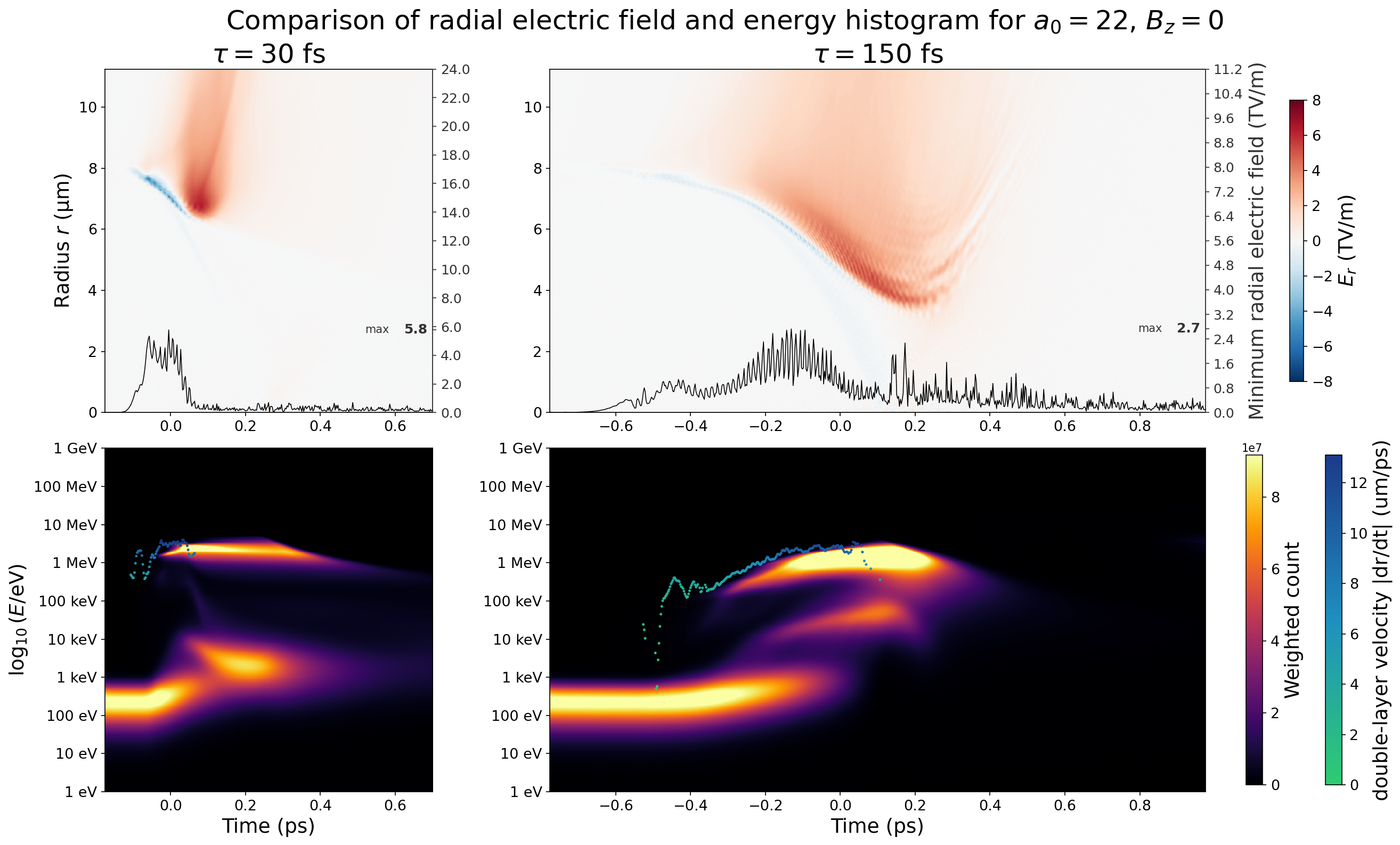}
\caption{Kinematic analysis of the 30~fs (left column) and 150~fs (right column) pulses at $a_0 = 22.0$. (Top) $E_r(r,t)$ colourmap; the dot markers tracking the front trajectory (shown at $a_0 = 12.7$ in Figure~\ref{fig:kinematics12p7}) are omitted here to avoid obscuring the field structure, while the black $-\min E_r$ trace and its annotated peak are retained. (Bottom) Inward-ion energy histograms with the $2v_{hb}$ reflection-energy markers overlaid (convention defined in the main text).}
\label{fig:kinematics22}
\end{adjustwidth}
\end{figure}

\begin{figure}[htbp]
\begin{adjustwidth}{-\extralength}{0cm}
  \centering
    \includegraphics[width=1\linewidth]{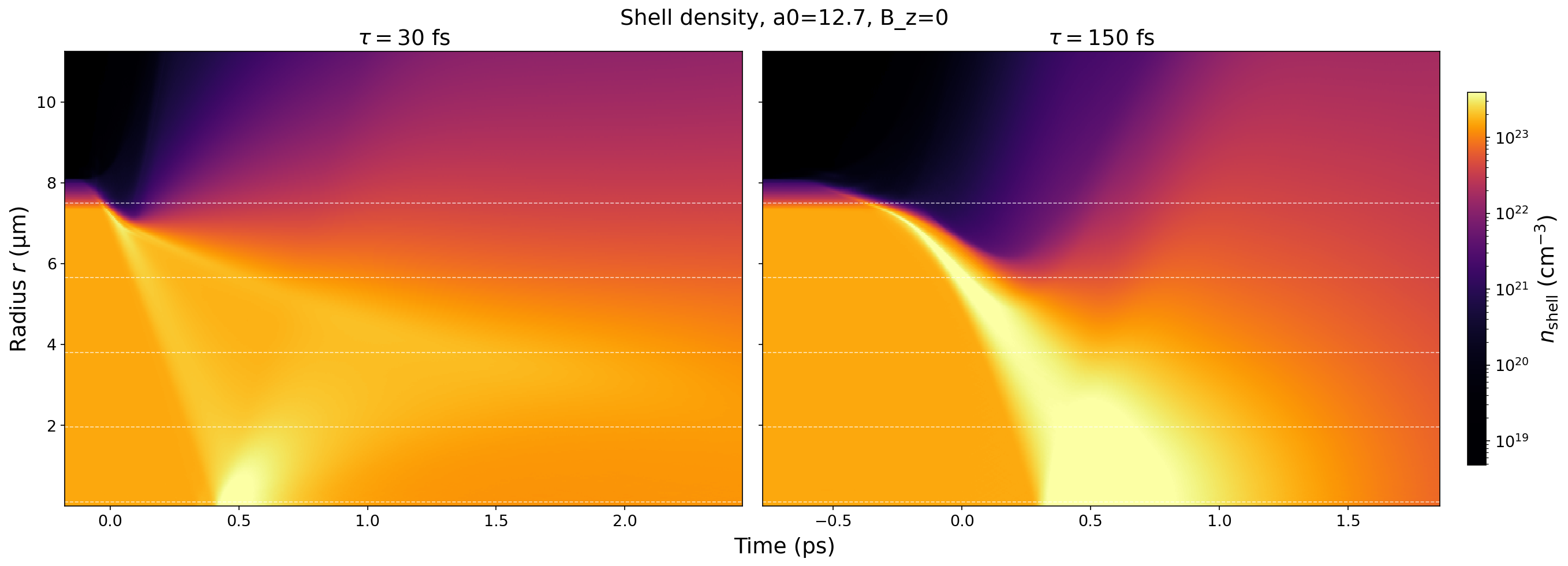}
  \caption{Shell density $n(r,t)$ in the radius--time plane (see main text for the definition of the radial shells and the cm$^{-3}$ normalisation) for the 30~fs (left) and 150~fs (right) pulses at $a_0 = 12.7$. The 30~fs case resolves two distinct inward-propagating density tracks, corresponding to the fast ($\sim 1$~MeV, steeper slope) and bulk ($10$--$100$~keV, shallower slope) ion populations. Under the 150~fs driver, the continuously accelerating charge-separation front smears these two populations into a single thick convergent feature. For the 30~fs driver the fast-ion ridge is approximately linear in $(r,t)$, consistent with a nearly constant propagation speed of the charge-separation front during the brief rising edge. For the 150~fs driver the ridge is curved because the electrostatic double-layer (hole-boring piston) accelerates throughout the extended rising edge, launching ions at progressively higher velocities; this same time-dependent launch energy appears in the histograms as a broadened, swept fast-ion band rather than a monoenergetic trace.}
  \label{fig:density_rt}
\end{adjustwidth}
\end{figure}

\begin{figure}[htbp]
\begin{adjustwidth}{-\extralength}{0cm}
  \centering
    \includegraphics[width=1\linewidth]{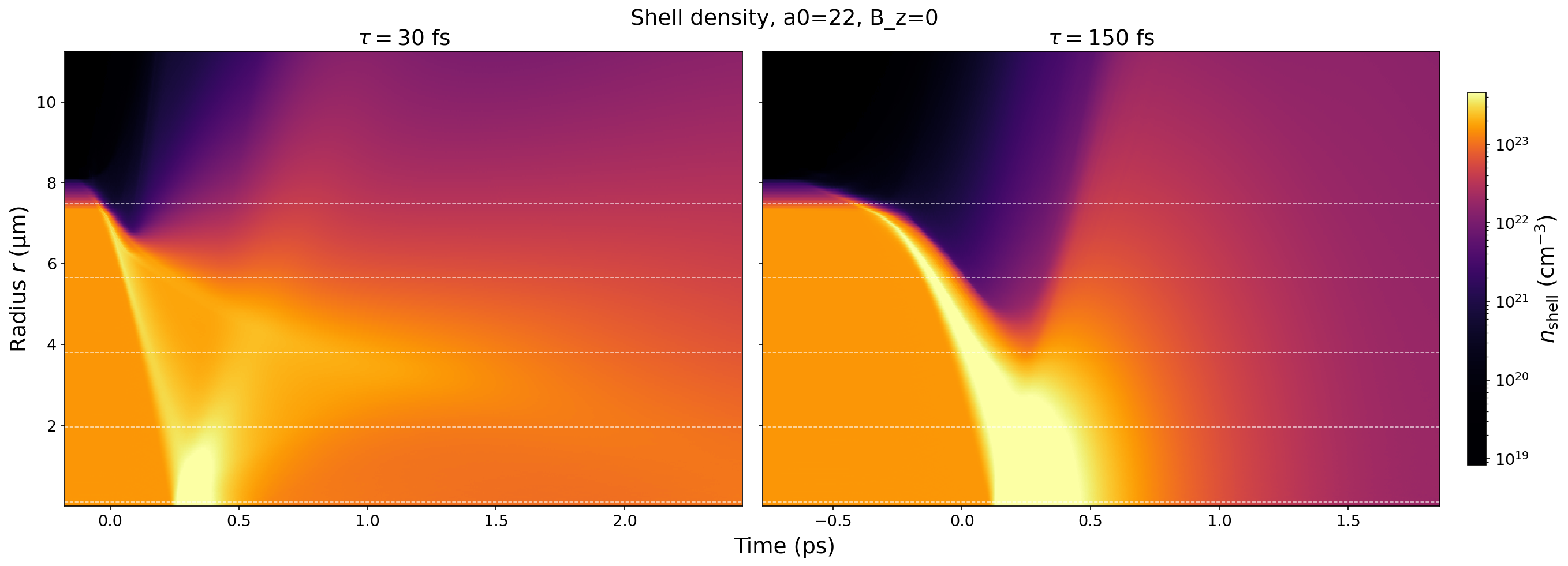}
  \caption{Shell density $n(r,t)$ in the radius--time plane (same definition as in Figure~\ref{fig:density_rt}) for the 30~fs (left) and 150~fs (right) pulses at the elevated intensity $a_0 = 22.0$. The qualitative picture is unchanged with respect to $a_0 = 12.7$: the 30~fs driver still resolves two distinct fast/bulk density tracks, while the 150~fs driver produces a single thick convergent feature. The tracks are, however, more sharply imprinted and steeper in the $(r,t)$ plane, consistent with the upshift of the fast-ion band into the $1$--$5$~MeV range.}
  \label{fig:density_rt_a22}
\end{adjustwidth}
\end{figure}

\begin{adjustwidth}{-\extralength}{0cm}

\subsection{The Non-Quasi-Neutral Charge-Separation Front}\label{sec:sound_speed_vs_dl}

The charge-separation front whose trajectory is extracted in the previous subsection is intrinsically non-quasi-neutral: it consists of a compressed electron sheet leading a depleted, positively charged ion cavity, sustaining a coherent radial electric field of $\sim$~TV/m amplitude over hundreds of femtoseconds (Figure~\ref{fig:electric_fields}). The fast-ion population is launched by direct reflection off this moving electrostatic structure.

This mechanism lies outside the closure assumptions of standard radiation-hydrodynamic codes. Rad-hydro models treat the plasma as a quasi-neutral, charge-balanced fluid with a single ion velocity field and an electron population represented only by a smooth temperature; the macroscopic electric field is enslaved to the pressure gradient through an Ohm- or ambipolar-type closure rather than being free to develop a coherent double-layer structure at the resolved scales of charge separation. By construction, such codes cannot represent a localized excess of negative charge driving inward at $\sim$~$10~\mu\mathrm{m}/\mathrm{ps}$ while a depleted positive shell lags behind, and they cannot host a separate, non-thermal fast-ion population reflected off that double-layer. The fast-ion channel resolved in Figures~\ref{fig:histograms}, \ref{fig:kinematics12p7}, and~\ref{fig:kinematics22} is therefore a genuinely kinetic feature of the interaction and requires a fully kinetic simulation to be captured.

\end{adjustwidth}
\begin{adjustwidth}{-\extralength}{0cm}

\subsection{Radial Density Evolution and Visualization of the Two Ion Populations}

The kinematic bifurcation inferred from the energy histograms (Figure~\ref{fig:histograms}) is directly reflected in the spatiotemporal evolution of the target density itself. Figure~\ref{fig:density_rt} presents the \emph{azimuthally averaged shell density} $n(r,t)$ in the radius--time plane for the 30~fs and 150~fs pulses at $a_0 = 12.7$. We use the term ``shell'' because, at each radius $r$ and time $t$, $n(r,t)$ is the mean particle number density evaluated inside a thin spherical shell of thickness $\Delta r_{\text{bin}}$ centred on $r$: the radial axis is built from 300 uniformly spaced bins spanning the range $r \in [0, L/2]$, where $L = 30.3~\mu\mathrm{m}$ is the in-plane domain size and the factor of two reflects that $r$ is measured from the target centre. This yields a radial bin width of $\Delta r_{\text{bin}} = (L/2)/300 \approx 50.5$~nm, i.e.\ roughly 17 native simulation cells. Within each shell, the per-area particle count is averaged over the polar angle and then converted to a per-volume density under the slab-depth convention of Section~\ref{sec:numerical_domain}.

Under the impulsive 30~fs driver (left panel), two distinct inward-propagating density features are clearly resolved. A steep, narrow ridge separates rapidly from the outer target surface and traverses the full target radius in a few hundred femtoseconds, tracing a nearly straight line in the $(r,t)$ plane. This corresponds to the fast, ballistic ion population reflected off the transient charge-separation front. A second, shallower trajectory lags behind at roughly half the inward velocity; it marks the slower hole-boring density wave carrying the bulk ions. The two features remain visually separated throughout the entire inward transit, providing a direct real-space signature of the monoenergetic $\sim 1$~MeV fast-ion band and the $10$--$100$~keV bulk population identified in the momentum histograms.

In contrast, the sustained 150~fs driver (right panel) does not exhibit a comparable clean separation. Because the charge-separation front continuously accelerates during the extended rising edge of the pulse, reflected fast ions are launched over a broad range of velocities, and the bulk hole-boring front is actively pushed deeper for several hundred femtoseconds. The two populations therefore overlap in the $(r,t)$ plane, blurring into a single, thick convergent feature. This real-space blurring is the direct counterpart of the non-monoenergetic energy sweep observed in Figure~\ref{fig:histograms} for the 150~fs case.

\end{adjustwidth}
\begin{adjustwidth}{-\extralength}{0cm}

Scaling the incident intensity to $a_0 = 22.0$ (Figure~\ref{fig:density_rt_a22} presents the corresponding maps) introduces no qualitative change in the density evolution: the same two-track structure for the 30~fs driver and the same single-feature signature for the 150~fs driver are recovered, in full consistency with the preserved kinematic bifurcation already established from the energy spectra (Figure~\ref{fig:kinematics22}). Quantitatively, however, the stronger ponderomotive drive produces a sharper, more contrast-rich imprint of both ion populations on the $n(r,t)$ plane. In the 30~fs case, the fast and bulk tracks are even more cleanly separated than at $a_0 = 12.7$, and both tracks are visibly steeper in the $(r,t)$ plane: their slopes $dr/dt$ correspond to inward speeds consistent with the $1$--$5$~MeV fast-ion band and the $\sim 10$~keV bulk, respectively. For the 150~fs driver, the single thick convergent feature similarly exhibits a shorter transit time, reflecting the higher mean inward velocity of the piston-driven population. 

\end{adjustwidth}
\begin{adjustwidth}{-\extralength}{0cm}

\section{Results Part II: Magnetic Field Effects}\label{sec:magnetic_results}

Following the baseline unmagnetized study, we evaluate the influence of an external axial magnetic field ($B_z$) on the laser-driven compression. The primary goal of applying this field is to constrain the transverse transport of laser-heated hot electrons, thereby mitigating energy loss and suppressing premature radial target expansion.

\subsection{The Lab-Achievable Regime (20 T)}\label{sec:20T}

We first consider the application of a 20~T external magnetic field, representing a magnitude that is readily achievable using modern pulsed-power coils in experimental laser facilities. Across all laser configurations (30~fs and 150~fs, at both $a_0 = 12.7$ and $a_0 = 22.0$), the introduction of the 20~T field produces no observable macroscopic difference compared to the unmagnetized baseline. The evolution of the target density, the structure of the charge-separation fronts, and the resulting ion energy spectra remain virtually identical to the $B_z = 0$ case. 

To provide a comprehensive view of the entire parameter space, including the identical diagnostic signatures of the 0~T and 20~T cases, an interactive repository of the simulation results has been made available online\cite{Optolowicz2026}.

The negligible impact of the 20~T field can be understood quantitatively through the relativistic Larmor radius,
\begin{equation}
  r_L(E_k, B_z, m) \;=\; \frac{p_\perp}{|q|\,B_z} \;=\; \frac{\sqrt{E_k\,(E_k + 2 m c^2)}}{|q|\,B_z\, c},
  \label{eq:larmor}
\end{equation}
where $p_\perp$ is the particle momentum component perpendicular to the magnetic field, $q$ is the particle charge (so $|q| = e$ for both electrons and protons), $m$ is the particle mass, and $c$ is the speed of light.
We evaluate the equation at a representative set of kinetic energies $E_k$ and species ($m = m_e$ or $m_p$) spanning the full range of particle populations observed in our simulations. Values across the four experimentally and theoretically relevant field strengths are collected in Table~\ref{tab:larmor}. At 20~T the gyro-radius for a typical MeV-class hot electron is several hundred micrometres---more than an order of magnitude larger than the 15~$\mu$m initial target diameter---so hot electrons escape the interaction region before completing even a fraction of a gyro-orbit. Protons are immune on an even wider margin, with $r_L$ in the millimetre range at every relevant energy. The 20~T field is therefore effectively invisible to the bulk expansion dynamics, exactly as the simulations confirm.

\end{adjustwidth}
\begin{table}[htbp]
\begin{adjustwidth}{-\extralength}{0cm}
  \centering
  \caption{Relativistic Larmor radius $r_L$ from Equation~\eqref{eq:larmor}, in micrometres, for electrons (e$^{-}$) and protons (p$^{+}$) of kinetic energy $E_k$ in an axial magnetic field $B_z$. In each cell is given Larmor radius for electrons and protons. Cells in bold indicate species whose gyro-radius is smaller than the target radius $R = 7.5~\mu$m and are therefore magnetised on the target scale. The values highlight the stark mass-induced asymmetry: electrons become magnetised already at $\sim 1$~kT (for $E_k \lesssim 1$~MeV), whereas protons only reach target-scale confinement at 10~kT and only for the lowest kinetic energies considered.}
  \label{tab:larmor}
  \setlength{\tabcolsep}{6pt}
  \renewcommand{\arraystretch}{1.25}
  \begin{tabular}{lccc}
    \toprule
    $B_z$ & $E_k = 100$~keV & $E_k = 1$~MeV & $E_k = 10$~MeV \\
     & (e$^{-}$ / p$^{+}$, $\mu$m) & (e$^{-}$ / p$^{+}$, $\mu$m) & (e$^{-}$ / p$^{+}$, $\mu$m) \\
    \midrule
    20~T    & 55.9 / 2280           & 237 / 7230           & 1750 / 22900 \\
    1~kT    & \textbf{1.12} / 45.7  & \textbf{4.74} / 145  & 35.0 / 458 \\
    5~kT    & \textbf{0.223} / 9.14 & \textbf{0.949} / 28.9 & \textbf{7.00} / 91.6 \\
    10~kT   & \textbf{0.112} / \textbf{4.57} & \textbf{0.474} / 14.5 & \textbf{3.50} / 45.8 \\
    \bottomrule
  \end{tabular}
\end{adjustwidth}
\end{table}
\begin{adjustwidth}{-\extralength}{0cm}

\subsection{Compression Time from the Inward-Momentum Ratio}\label{sec:electron_dynamics}

Because the macroscopic response of the target to magnetisation can be quantified by how long the system spends with a net inward particle flux, we build a time-resolved diagnostic from the full kinetic-energy histograms. Starting from the inward-momentum and total histograms introduced in Section~\ref{sec:energy_Histograms_introduction}, we compute the energy-resolved quantity
\begin{equation}
  R(E_k,t) \;=\; \frac{W_{\text{inward}}(E_k,t)}{W_{\text{all}}(E_k,t)} \;-\; \frac{1}{2},
  \label{eq:R_map}
\end{equation}
where $W_{\text{inward}}$ and $W_{\text{all}}$ are the total weighted macroparticle counts summed over all simulated species in each $(E_k,t)$ bin---both electrons and ions. By construction, $R = 0$ means that particles at that kinetic energy are moving inward and outward in equal proportion (no net radial flux), $R > 0$ flags an excess of inward-directed particles, and $R < 0$ an excess of outward-directed ones. A visualization of this quantity is realized in Figure~\ref{fig:inward_ratio_2d}.
The all-species integral is the natural choice here because the compression versus blow-off transition of the target is a bulk dynamical event in which electrons and ions both participate, and the two species populate complementary regions of the $(E_k,t)$ plane. An electron-only version of the same diagnostic yields a qualitatively different picture, which we return to at the end of this subsection.

\end{adjustwidth}
\begin{figure}[htbp]
\begin{adjustwidth}{-\extralength}{0cm}
  \centering
  \includegraphics[width=0.95\linewidth]{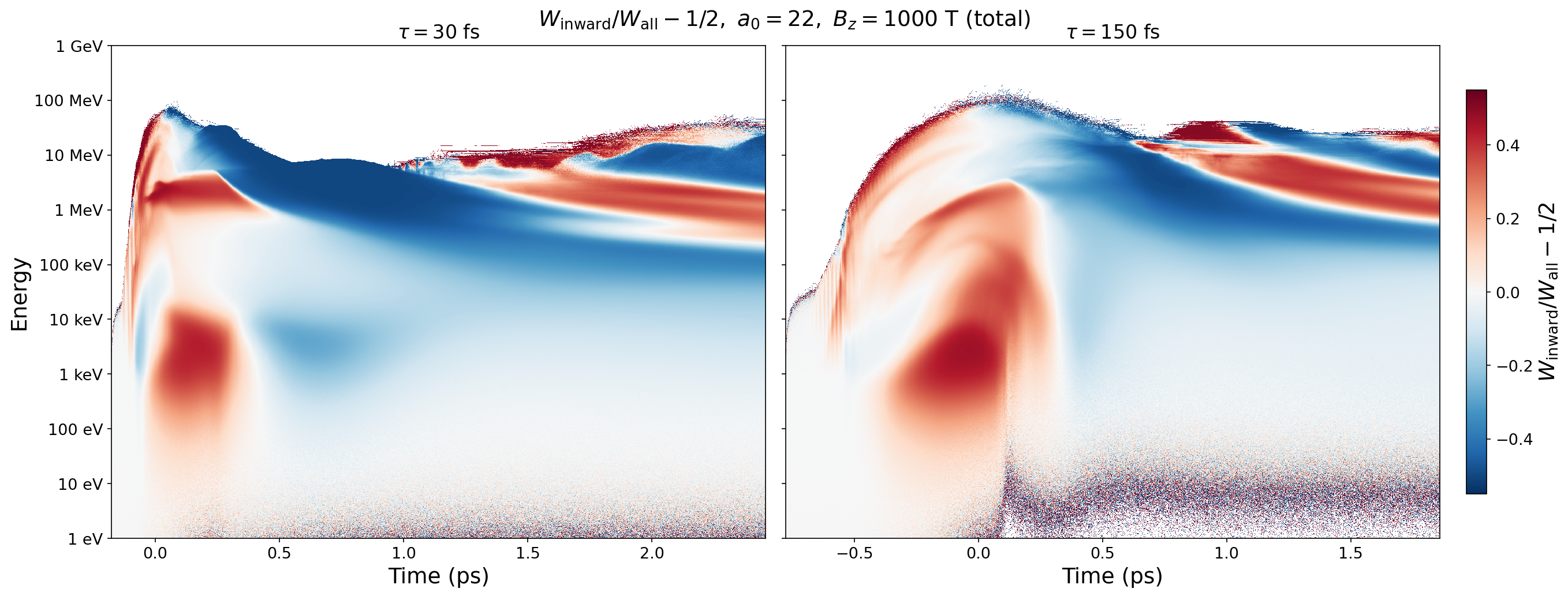}
  \caption{Energy-resolved inward-fraction map at $a_0 = 22.0$ and $B_z = 1$~kT, for $\tau = 30$~fs (left) and $\tau = 150$~fs (right). The $E_k$ axis is identical to the kinetic-energy axis used in the energy histograms; only the colour-encoded quantity differs. Red (positive) pixels indicate populations whose radial momentum is predominantly inward-directed at that kinetic energy and time; blue (negative) pixels indicate predominantly outward-directed populations. The map aggregates over all simulated species (electrons and H$^{+}$). A single representative $(\tau, a_0, B_z)$ combination is shown here for illustration; the full set of maps for every simulated configuration is available interactively at \url{https://sim-results.netlify.app/}~\cite{Optolowicz2026}. We do not attempt to attribute individual spectral features of these maps to specific physical processes in this work; they are introduced as the raw input from which we derive the integrated diagnostics of Figures~\ref{fig:inward_ratio_1d} and \ref{fig:t_comp_summary}.}
  \label{fig:inward_ratio_2d}
\end{adjustwidth}
\end{figure}
\begin{adjustwidth}{-\extralength}{0cm}

To condense this 2D information into a time series, we integrate $R$ over kinetic energy at each timestep, obtaining a single inward-fraction curve
\begin{equation}
  r(t) \;=\; \frac{\int W_{\text{inward}}(E_k,t)\,\mathrm{d}E_k}{\int W_{\text{all}}(E_k,t)\,\mathrm{d}E_k} \;-\; \frac{1}{2}.
  \label{eq:r_curve}
\end{equation}
Examples are shown in Figure~\ref{fig:inward_ratio_1d}: when $r(t) > 0$ the system carries a net inward particle flux, and when $r(t) < 0$ it is in the ablation / blow-off phase. We define the \emph{compression time} $t_{\text{comp}}$ as the length of the longest contiguous interval for which $r(t) > 0$ (shaded region in Figure~\ref{fig:inward_ratio_1d}). In the unmagnetized case, $r(t)$ exhibits a brief dip below zero near the leading edge of the pulse before recovering into the main inward-flux interval; this initial excursion becomes progressively less pronounced as $B_z$ is increased and is no longer resolved for $B_z \gtrsim 5$~kT, as visible directly in the 1$\times$4 field scan of Figure~\ref{fig:inward_ratio_1d}. We do not claim a specific microphysical mechanism for the suppression of this initial dip, since the all-species ratio mixes electron- and ion-driven contributions; we note only that its disappearance with increasing $B_z$ is consistent with a more laminar, less ablation-dominated leading edge of the interaction.

\end{adjustwidth}
\begin{figure}[htbp]
\begin{adjustwidth}{-\extralength}{0cm}
  \centering
  \begin{tabular}{@{}c@{}}
    \includegraphics[width=0.6\linewidth]{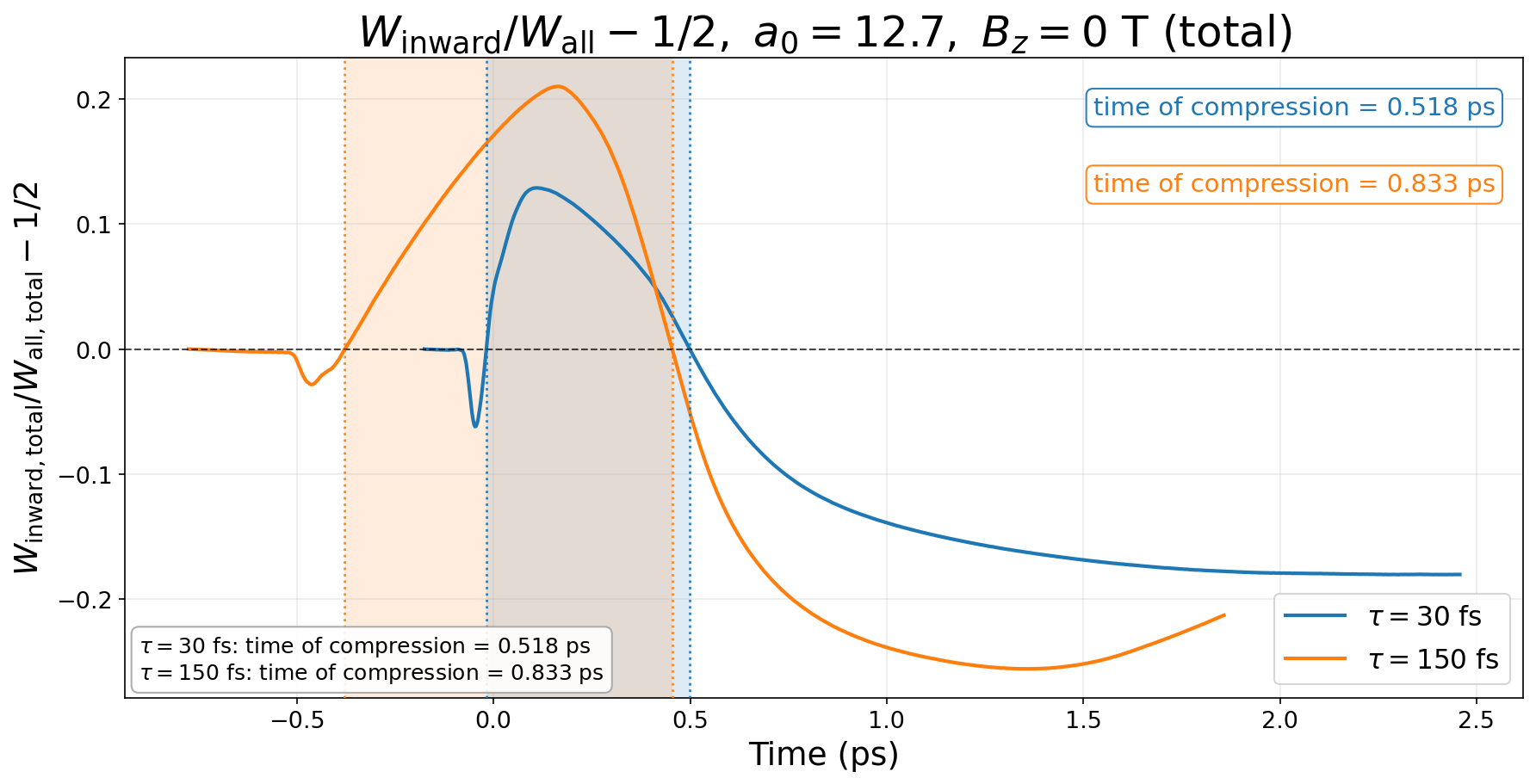}\\
    \includegraphics[width=0.6\linewidth]{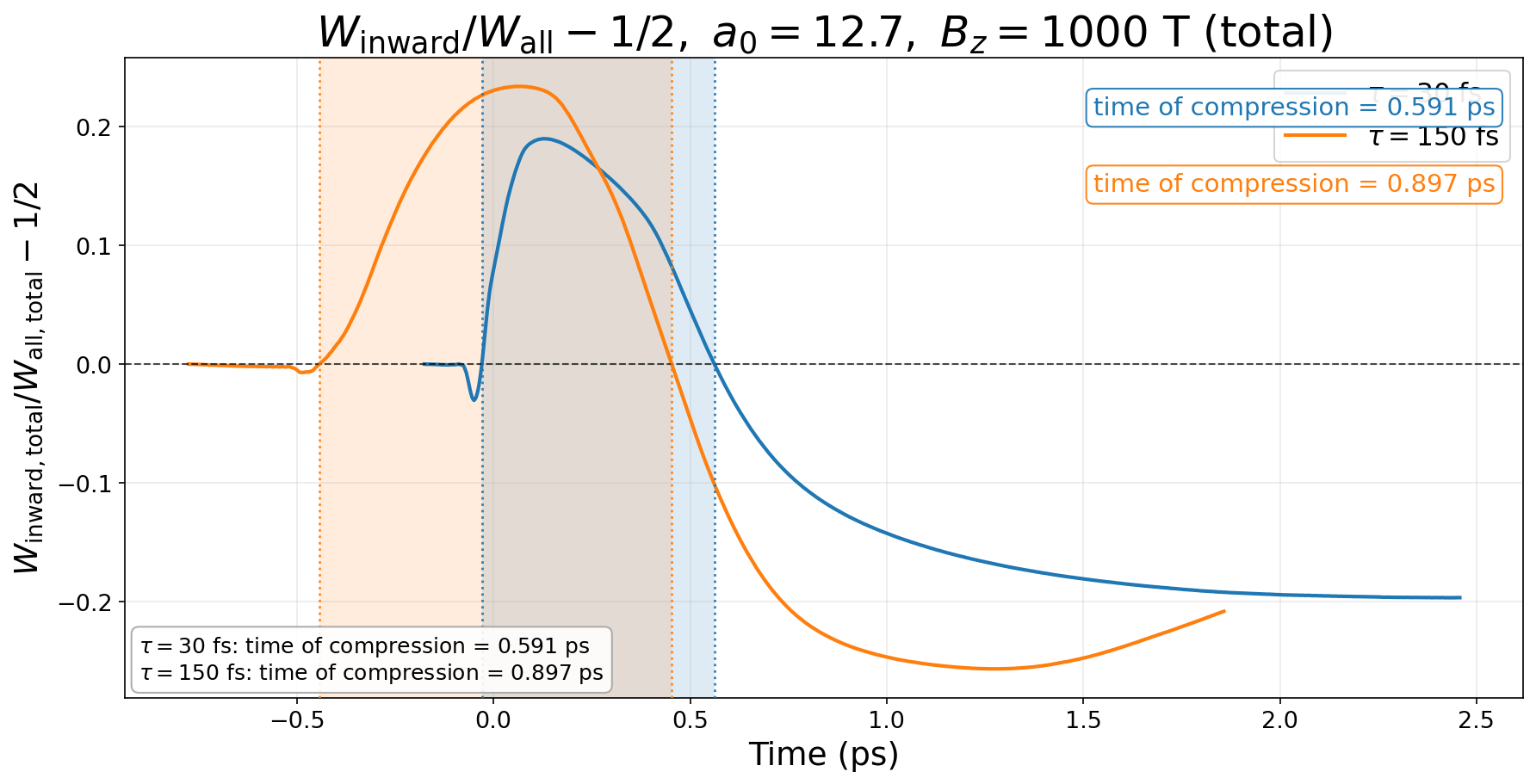}\\
    \includegraphics[width=0.6\linewidth]{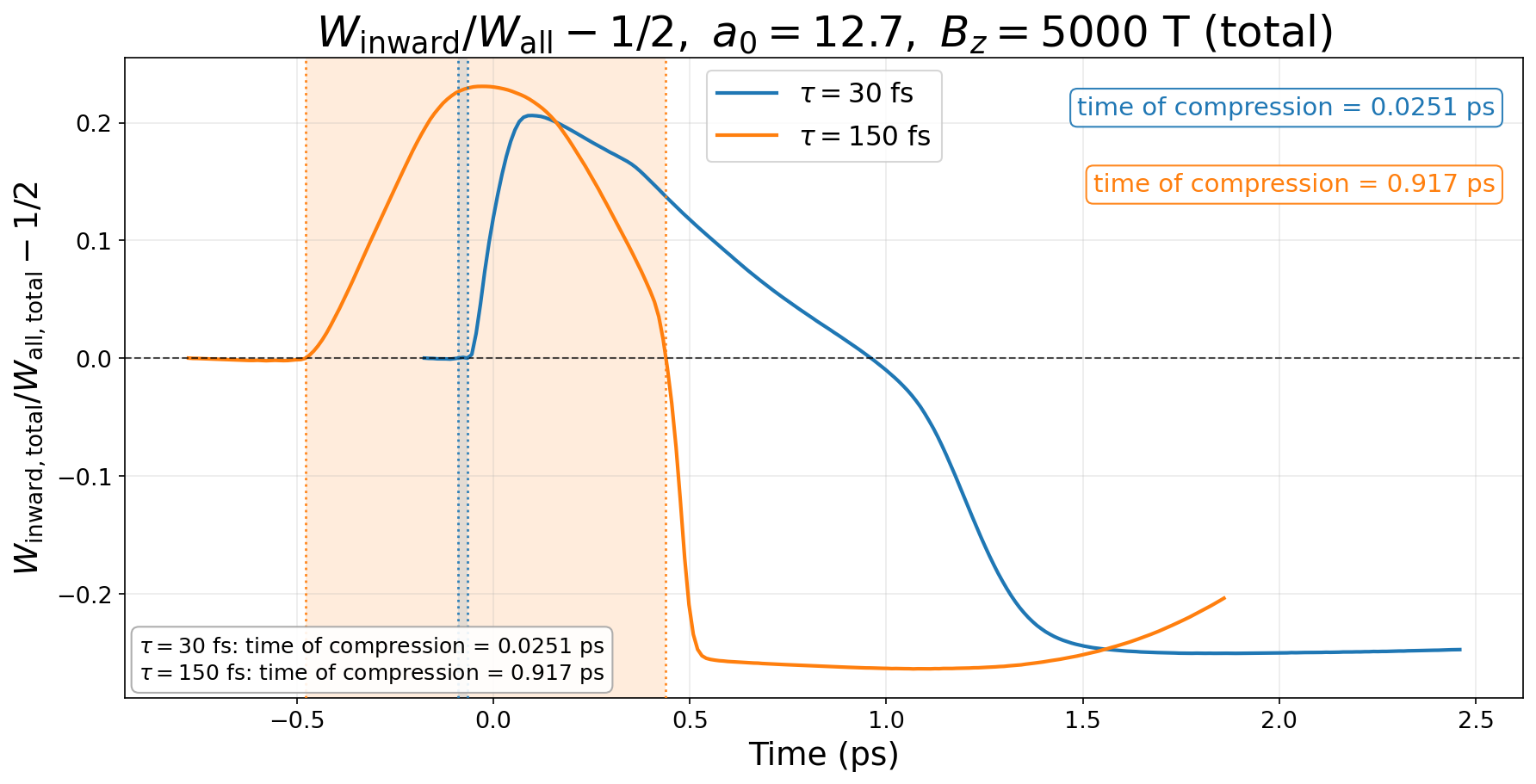}\\
    \includegraphics[width=0.6\linewidth]{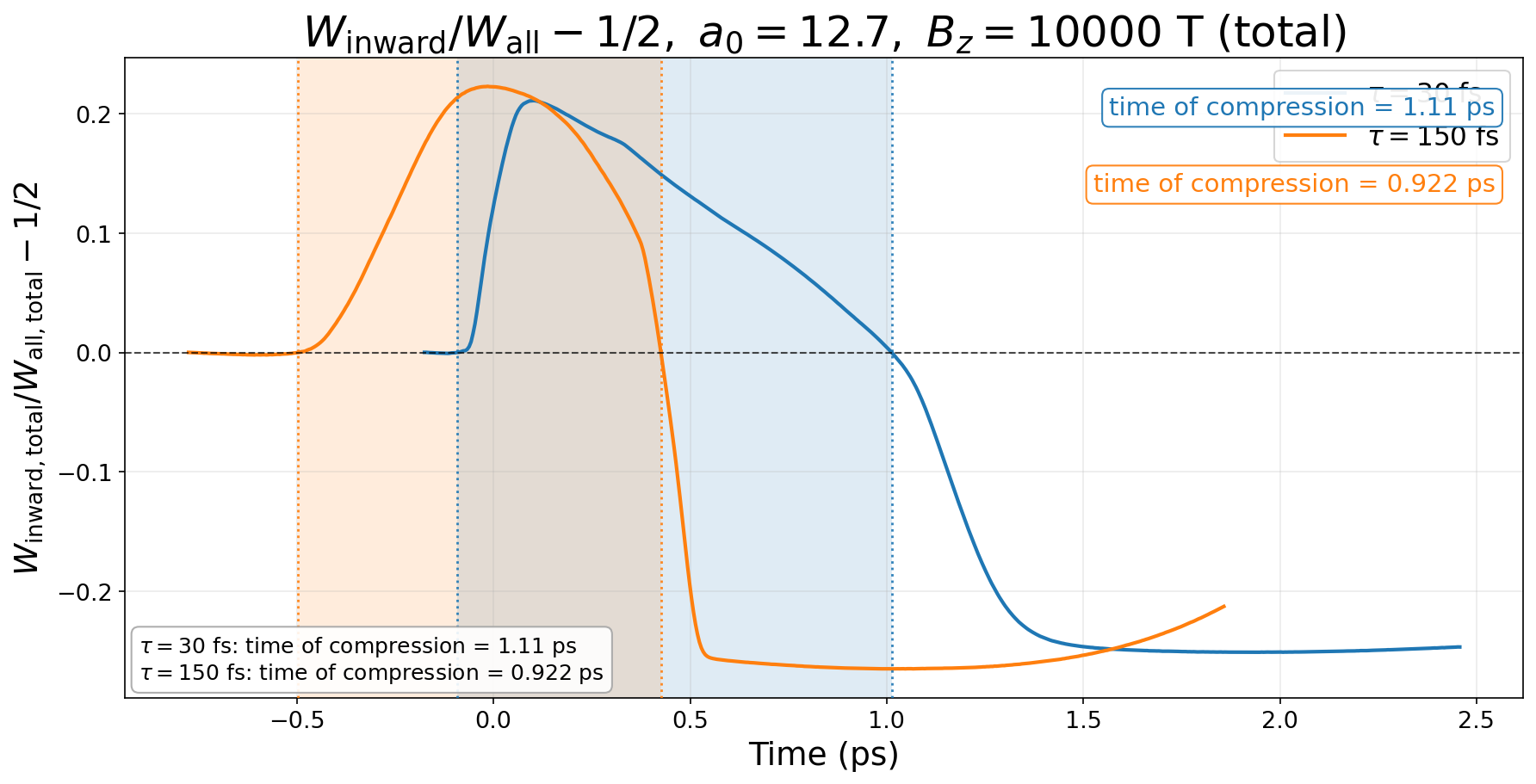}\\
  \end{tabular}
  \caption{Energy-integrated inward-fraction curve $r(t)$ from Equation~\eqref{eq:r_curve}, shown here at $a_0 = 12.7$ for both pulse durations, for four external fields $B_z \in \{0, 1, 5, 10\}$~kT. The shaded regions mark the longest contiguous interval over which $r(t) > 0$; its length defines the compression time $t_{\text{comp}}$ annotated for each driver. The brief early-time negative excursion visible at $B_z = 0$ becomes progressively less pronounced as $B_z$ is increased and is no longer resolved for $B_z \gtrsim 5$~kT. The full set of curves is available via the interactive viewer at \url{https://sim-results.netlify.app/}~\cite{Optolowicz2026}.}
  \label{fig:inward_ratio_1d}
\end{adjustwidth}
\end{figure}
\begin{adjustwidth}{-\extralength}{0cm}

Collecting $t_{\text{comp}}$ across the full parameter scan yields the summary of Figure~\ref{fig:t_comp_summary}, one panel per intensity, one curve per pulse duration. The headline trend is a clear separation between the two drivers. The 30~fs driver shows a strong, monotonic lengthening of the compression phase with $B_z$: at $a_0 = 12.7$, $t_{\text{comp}}$ more than doubles between $B_z = 0$ and $10$~kT (from $\approx 0.52$~ps to $\approx 1.1$~ps), and at $a_0 = 22.0$ it roughly doubles from $\approx 0.36$~ps to $\approx 0.73$~ps. The 150~fs driver shows a much weaker response: its $t_{\text{comp}}$ is already long in the unmagnetized case ($\approx 0.84$~ps at $a_0 = 12.7$ and $\approx 0.67$~ps at $a_0 = 22.0$) and grows only modestly with $B_z$, saturating around $\approx 0.92$~ps and $\approx 0.76$~ps respectively.

This asymmetric response is consistent with the driver hierarchy established earlier. The 30~fs driver delivers an impulsive kick that, in the unmagnetized case, is quickly equilibrated; under strong axial magnetisation the net-inward phase of the target persists considerably longer, suggesting that the mechanisms which would otherwise restore the target to a net-outward blow-off state are suppressed in the magnetised regime. The 150~fs driver already sustains an inward push for most of its own duration, so magnetisation has less room to further lengthen the compression phase. Somewhat puzzling, at $a_0 = 12.7$ the two curves cross near $B_z \approx 4$~kT and the 30~fs driver produces a longer compression phase than the 150~fs driver at $B_z = 10$~kT---a reversal of the unmagnetized ordering. There is also a strong trend in the $a_0 = 22.0$ case that would point to this reversal at even higher magnetic fields.

\end{adjustwidth}
\begin{figure}[htbp]
\begin{adjustwidth}{-\extralength}{0cm}
  \centering
  \includegraphics[width=0.85\linewidth]{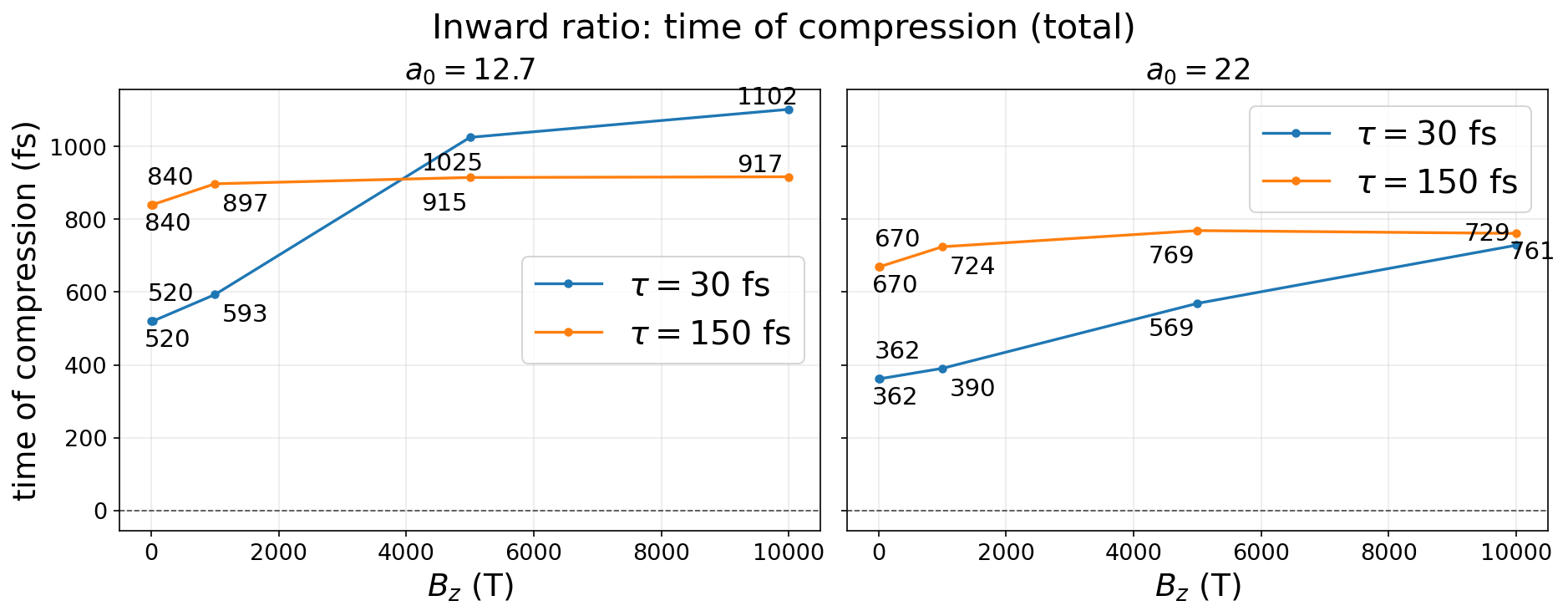}
  \caption{Compression time $t_{\text{comp}}$, defined as the longest contiguous interval over which the energy-integrated inward-fraction $r(t)$ of Equation~\eqref{eq:r_curve} remains positive, as a function of $B_z$ for each of the four $(\tau, a_0)$ configurations. Left panel: $a_0 = 12.7$; right panel: $a_0 = 22.0$. Blue curves: $\tau = 30$~fs; orange curves: $\tau = 150$~fs. The 30~fs driver shows a strong, monotonic lengthening of the compression phase with magnetisation, whereas the 150~fs driver is only weakly affected; at $a_0 = 12.7$ the two curves even cross near $B_z \approx 4$~kT.}
  \label{fig:t_comp_summary}
\end{adjustwidth}
\end{figure}
\begin{adjustwidth}{-\extralength}{0cm}

For completeness, Figure~\ref{fig:t_comp_summary_electrons} repeats the same compression-time diagnostic but restricted to the electron macroparticles only (i.e.\ $W_{\text{inward}}$ and $W_{\text{all}}$ in Equations~\eqref{eq:R_map}--\eqref{eq:r_curve} are computed from electrons alone). The picture that emerges is qualitatively distinct from the all-species case. At $a_0 = 12.7$ both drivers are non-monotonic: $t_{\text{comp}}^{\mathrm{e}^-}$ dips slightly for $B_z \in [1, 5]$~kT and rises at $10$~kT (from $472$~fs at $0$~T to $708$~fs at $10$~kT for the 30~fs driver). At $a_0 = 22.0$, by contrast, the 30~fs electron-only curve \emph{decreases} monotonically with $B_z$ (from $286$~fs at $0$~T to $261$~fs at $10$~kT), opposite to the all-species trend of Figure~\ref{fig:t_comp_summary}.

The electron-only curve measures how long the electron population is, on average, moving inward; the all-species curve measures how long the target as a whole is in its compression phase. The latter is the relevant quantity for assessing the duration of the inertial compression event, and it is the diagnostic we use in the rest of the section. The electron-only trend at $a_0 = 22.0$---a slight \emph{decrease} of $t_{\text{comp}}^{\mathrm{e}^-}$ with $B_z$ even though the all-species $t_{\text{comp}}$ \emph{increases}---is consistent with the picture developed in Section~\ref{sec:blowoff}: under strong magnetisation the electron population is locked into tight gyromotion, so its net-radial signature dissipates faster, while the resulting hot magnetised sheath continues to push the bulk (ion) target inward through the ambipolar field.

\end{adjustwidth}
\begin{figure}[htbp]
\begin{adjustwidth}{-\extralength}{0cm}
  \centering
  \includegraphics[width=0.85\linewidth]{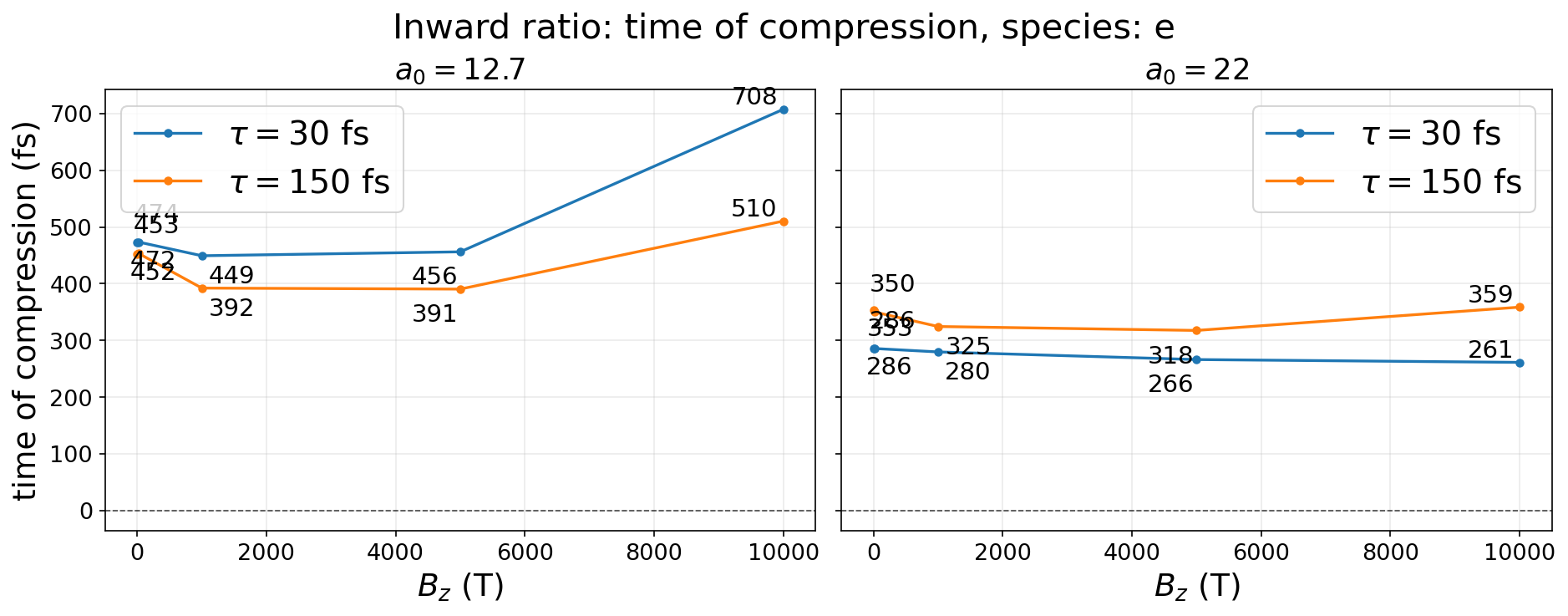}
  \caption{Electron-only variant of the compression time, obtained by restricting $W_{\text{inward}}$ and $W_{\text{all}}$ in Equations~\eqref{eq:R_map}--\eqref{eq:r_curve} to the electron macroparticles. Left panel: $a_0 = 12.7$; right panel: $a_0 = 22.0$. The resulting trends differ qualitatively from the all-species version in Figure~\ref{fig:t_comp_summary} and, in the $a_0 = 22.0$ / $\tau = 30$~fs case, have the opposite sign. This contrast motivates the discussion of the unexpected outward-extension effect in the main text.}
  \label{fig:t_comp_summary_electrons}
\end{adjustwidth}
\end{figure}
\begin{adjustwidth}{-\extralength}{0cm}

\subsection{Peak Compression Ratio at $r=1~\mu$m Across the Parameter Scan}\label{sec:compression_ratio_map}
The compression time $t_{\text{comp}}$ introduced above quantifies how long the target spends in its net-inward phase. To complement this duration metric with a measure of how much compression is achieved, we define a local compression ratio at a fixed radius $r = 1~\mu$m,
\begin{equation}
  C(t) \;=\; \frac{n(r=1~\mu\mathrm{m},t)}{n_0},
  \label{eq:compression_ratio}
\end{equation}
where $n(r,t)$ is the azimuthally averaged density and $n_0$ is the initial neutral density (Section~\ref{sec:numerical_domain}). The choice of $r = 1~\mu$m is somewhat arbitrary, but it strikes a practical balance: it is small enough to probe the compressed core, while still averaging over a sufficient number of cells to suppress bin-level noise. We summarize each run by its peak value $C_{\max} = \max_t C(t)$ (Figure~\ref{fig:compression_ratio_time}) and compare it across the full $(\tau, a_0, B_z)$ scan (Figure~\ref{fig:compression_ratio_heatmap}).

The $C(t)$ curves for the short, 30~fs driver (Figure~\ref{fig:compression_ratio_time}, top row) reveal a double-peak structure that is directly explained by the kinematic bifurcation established in Section~\ref{sec:energy_Histograms_introduction}. The early peak ($\sim 0.5$--$0.7$~ps) is produced by the fast, ballistic $\sim 1$~MeV ion population that is reflected off the charge-separation front and converges at the target center. Its timing and height are largely field-independent at low $B_z$ ($\sim 1$~kT), because the fast-ion channel remains open; the peak progressively shrinks and eventually disappears at kT-scale fields, where that channel is suppressed (Section~\ref{sec:ion_suppression}). The late peak (only visible at $B_z \gtrsim 1$~kT) is produced by the slower bulk hole-boring wave. In the unmagnetized case this wave decelerates once the impulsive 30~fs kick ends and ablation takes over, never converging fully at the core. Under strong magnetisation, however, $t_{\text{comp}}$ more than doubles (Figure~\ref{fig:t_comp_summary}), sustaining the inward push long enough for the bulk wave to reach $r=1~\mu$m and produce a second, larger density peak. The double-peak structure is therefore a manifestation of the bifurcation into fast and bulk ion populations, modulated by the magnetic suppression of the fast-ion channel and the extension of the bulk compression phase.

The 150~fs driver (bottom row) shows no comparable double peak: its sustained piston already drives the bulk all the way to the core during the pulse, producing a single peak that grows monotonically with $B_z$, consistent with its modest $t_{\text{comp}}$ response seen in Figure~\ref{fig:t_comp_summary}. The heatmap of Figure~\ref{fig:compression_ratio_heatmap} condenses $C_{\max}$ across the full scan; the strong rise at $B_z = 10$~kT in the 30~fs column is driven entirely by the late bulk peak, which has no counterpart in the unmagnetized case.

\end{adjustwidth}
\begin{figure}[htbp]
\begin{adjustwidth}{-\extralength}{0cm}
  \centering
  \includegraphics[width=0.9\linewidth]{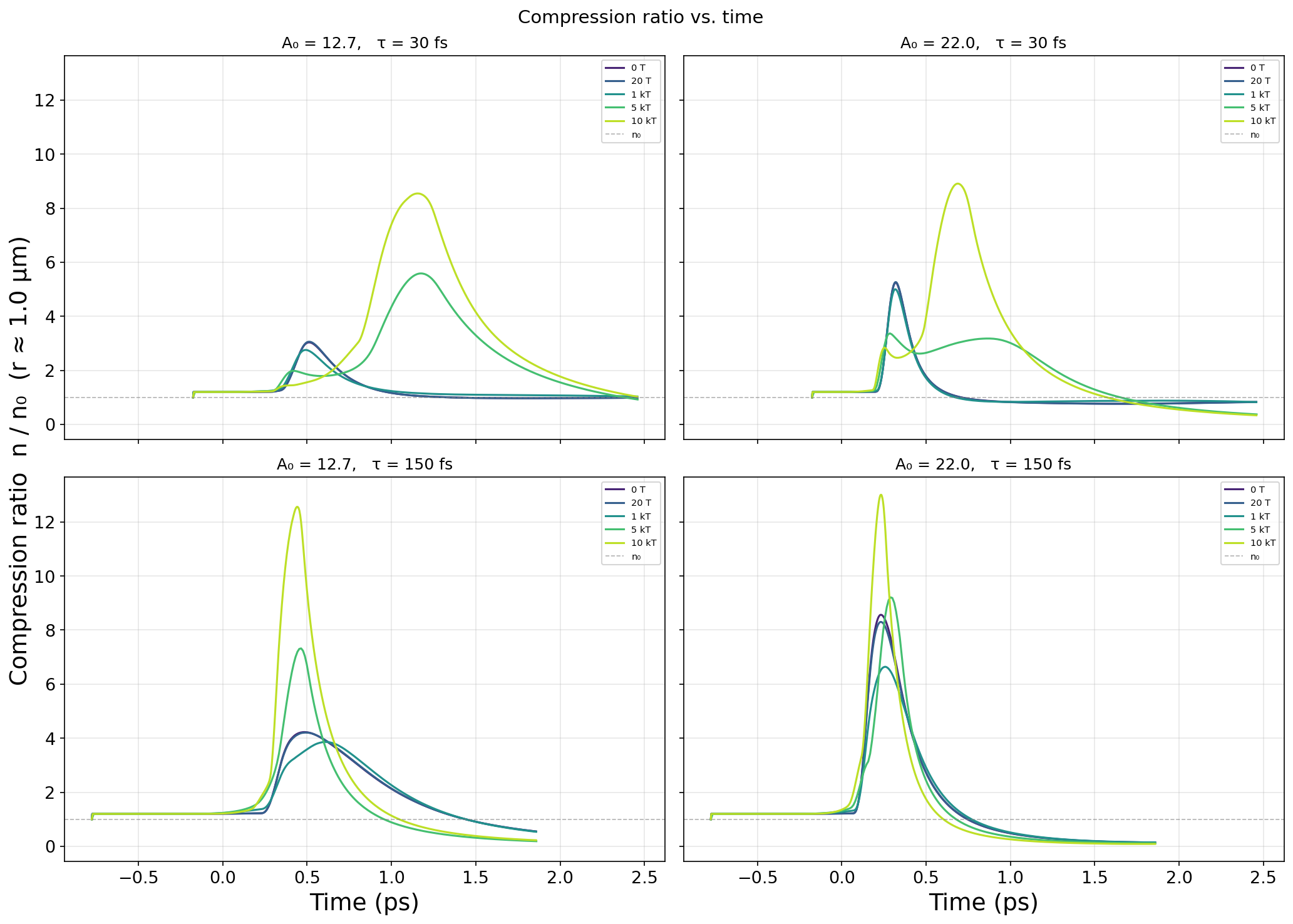}
  \caption{Time evolution of the local compression ratio $C(t) = n(r=1~\mu\mathrm{m},t)/n_0$ (Equation~\eqref{eq:compression_ratio}) for the four $(\tau, a_0)$ configurations of this work, with one curve per external field $B_z \in \{0, 1, 5, 10\}$~kT. Top row: $\tau = 30$~fs driver; bottom row: $\tau = 150$~fs driver. The 30~fs driver develops a double-peak structure at kT-scale fields (Section~\ref{sec:compression_ratio_map}) that is absent in the 150~fs case.}
  \label{fig:compression_ratio_time}
\end{adjustwidth}
\end{figure}

\begin{figure}[htbp]
\begin{adjustwidth}{-\extralength}{0cm}
  \centering
  \includegraphics[width=0.95\linewidth]{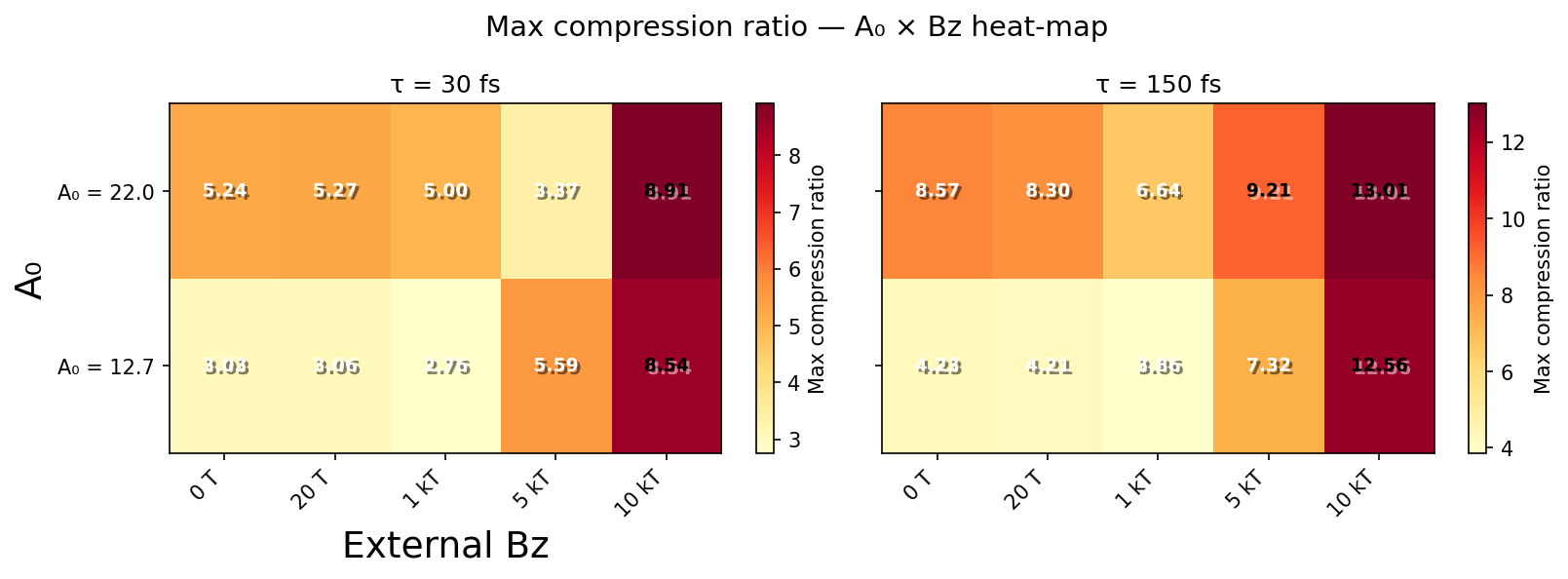}
  \caption{Peak compression ratio $C_{\max}=\max_t n(r=1~\mu\mathrm{m},t)/n_0$ across the full parameter scan, shown separately for $\tau=30$~fs (left) and $\tau=150$~fs (right). Cell annotations report $C_{\max}$ for each $(a_0,B_z)$ combination. This metric complements the compression time $t_{\text{comp}}$ of Figure~\ref{fig:t_comp_summary}: $t_{\text{comp}}$ measures how long the target remains in a net-inward phase, while $C_{\max}$ measures the maximum core-adjacent density reached during that phase.}
  \label{fig:compression_ratio_heatmap}
\end{adjustwidth}
\end{figure}
\begin{figure}[htbp]
\begin{adjustwidth}{-\extralength}{0cm}
  \centering
  \includegraphics[width=0.48\linewidth]{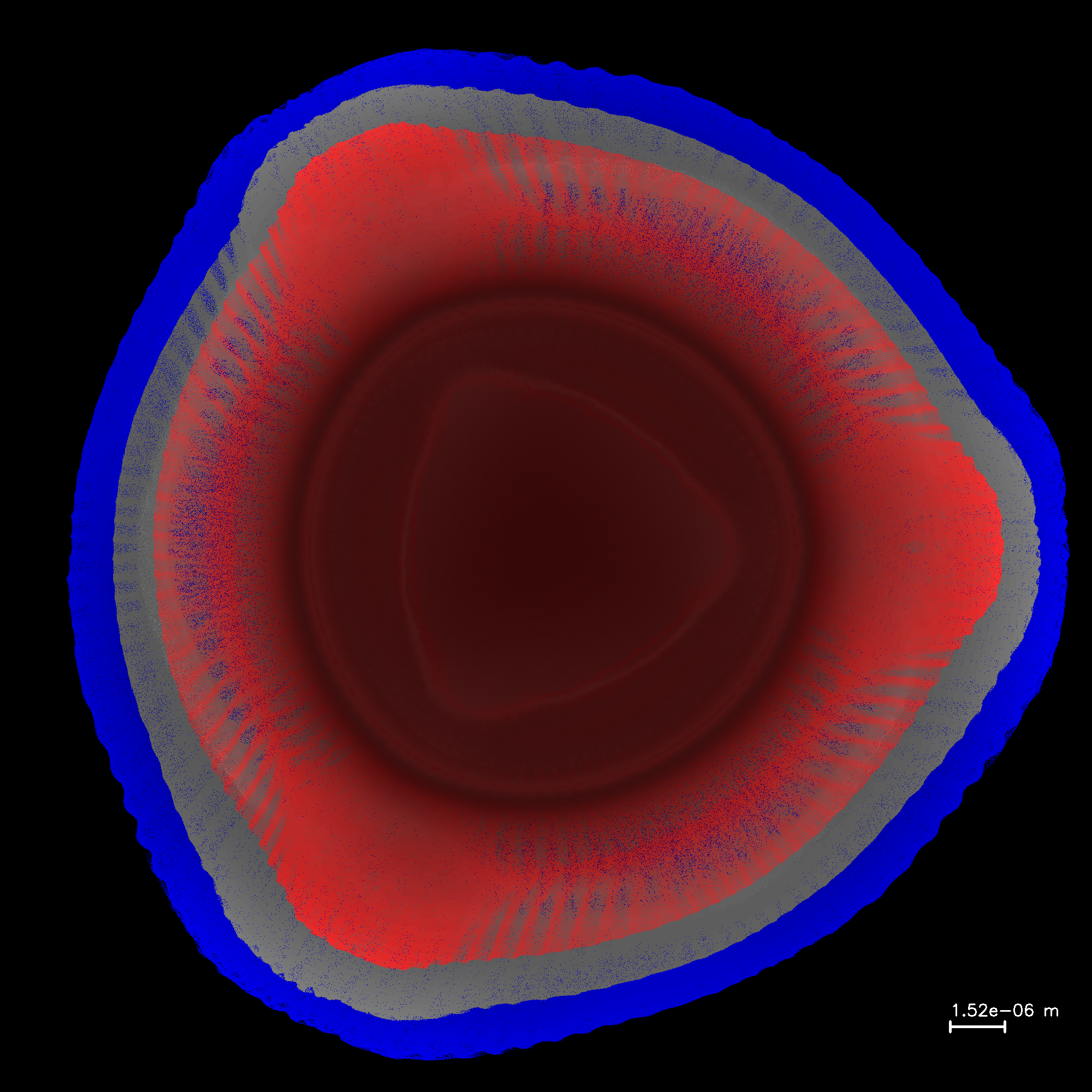}
  \includegraphics[width=0.48\linewidth]{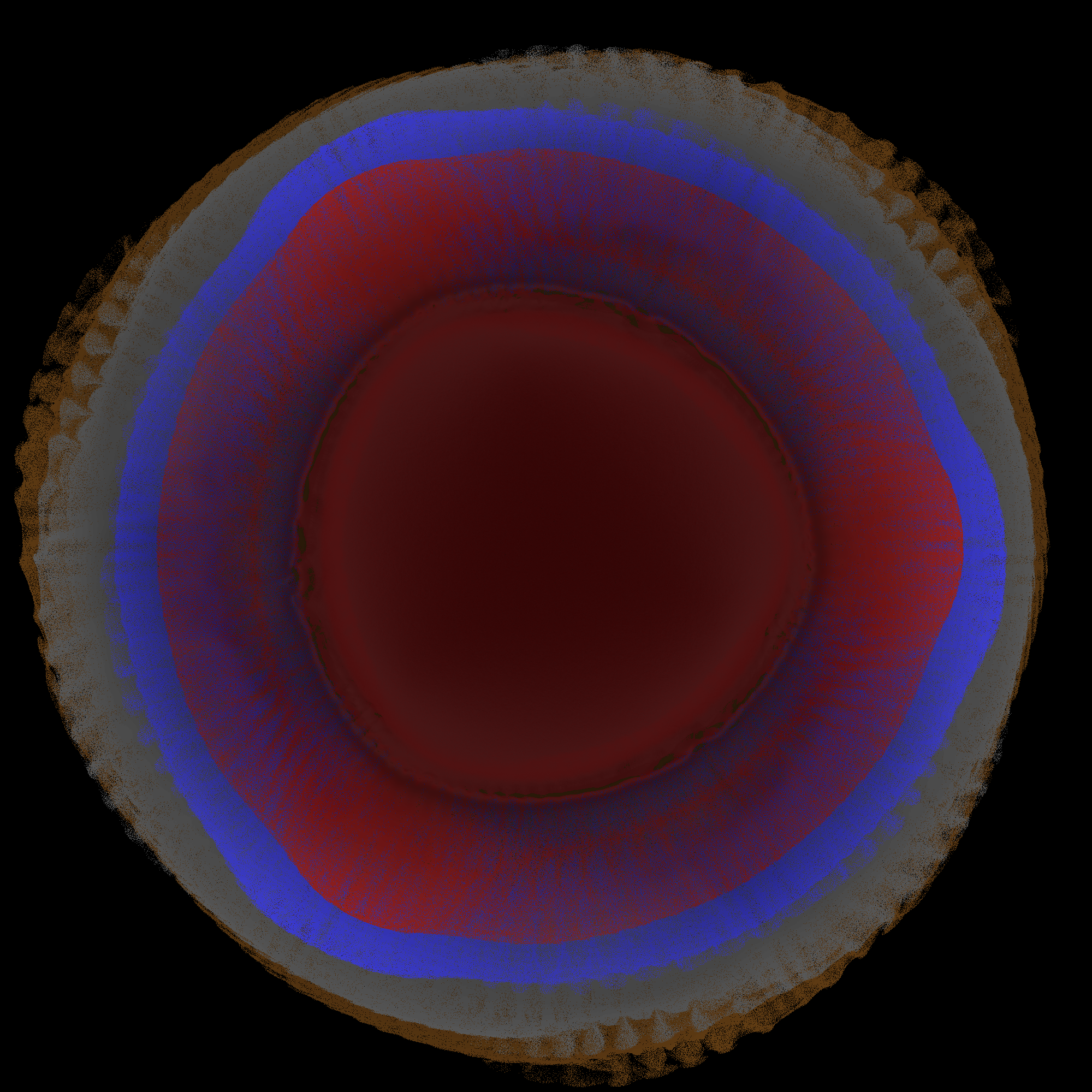}
  \caption{Overlaid ion kinetic-energy-density maps of the same target at the same time (before core collapse), for several values of the external axial magnetic field $B_z$, at $a_0 = 12.7$. Each overlay shows the outer iso-surface of the ion kinetic-energy density for one $B_z$ value, with the colour identifying $B_z$. Left: 30~fs driver, three overlays (red $= 0$~T, grey $= 5$~kT, blue $= 10$~kT). Right: 150~fs driver, four overlays (red $= 0$~T, blue $= 1$~kT, white/grey $= 5$~kT, golden $= 10$~kT). In both panels the envelope at higher $B_z$ extends \emph{further} from the target centre than the unmagnetized one, contrary to a naive single-particle confinement picture; the effect is more pronounced and the envelope separations are larger for the 30~fs driver. The same ordering is reproduced for the electron envelope (not shown).}
  \label{fig:envelope_overlay}
\end{adjustwidth}
\end{figure}
\begin{figure}[htbp]
\begin{adjustwidth}{-\extralength}{0cm}
  \centering

  \includegraphics[width=0.7\linewidth]{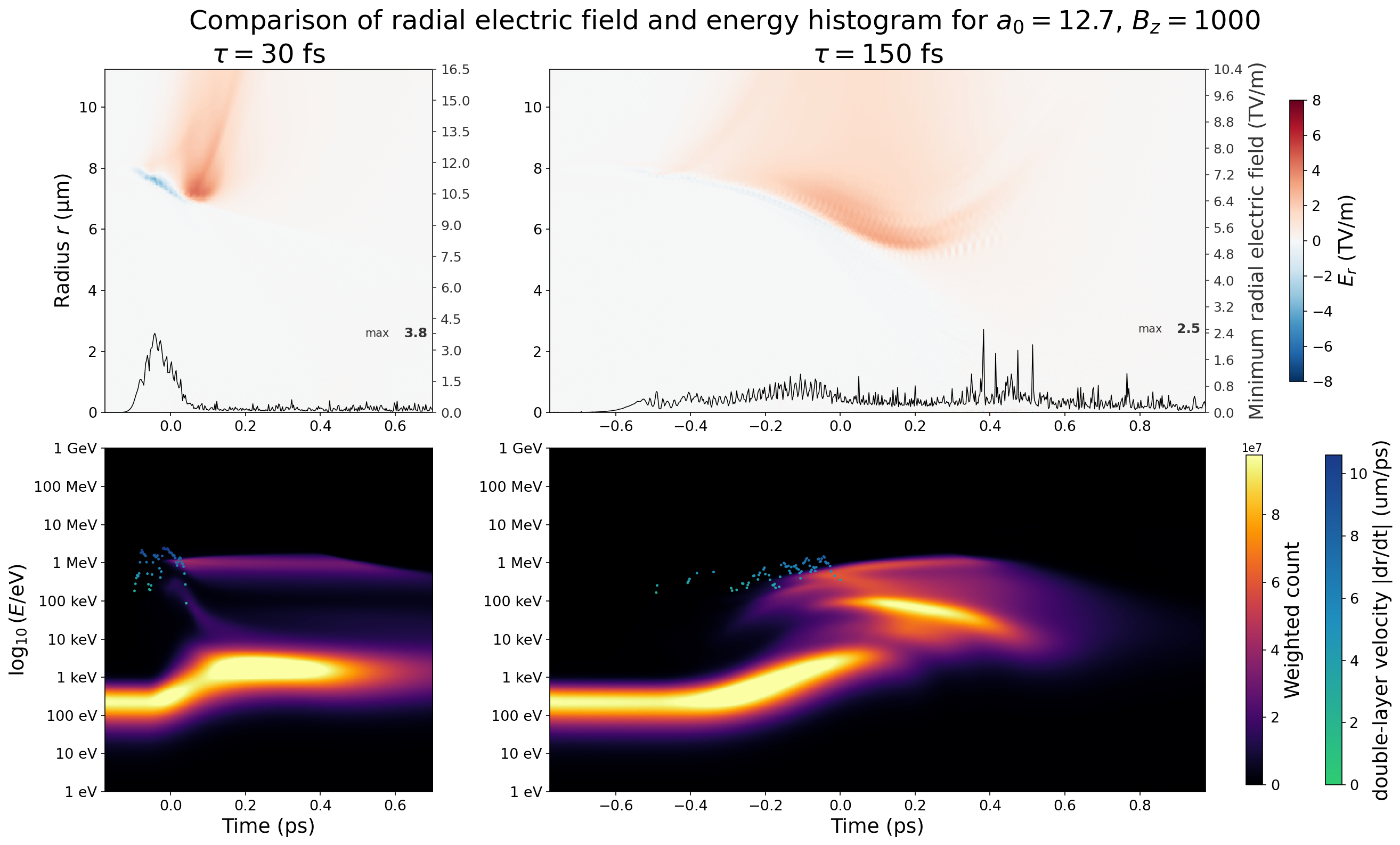}

  \vspace{0pt}

  \includegraphics[width=0.7\linewidth]{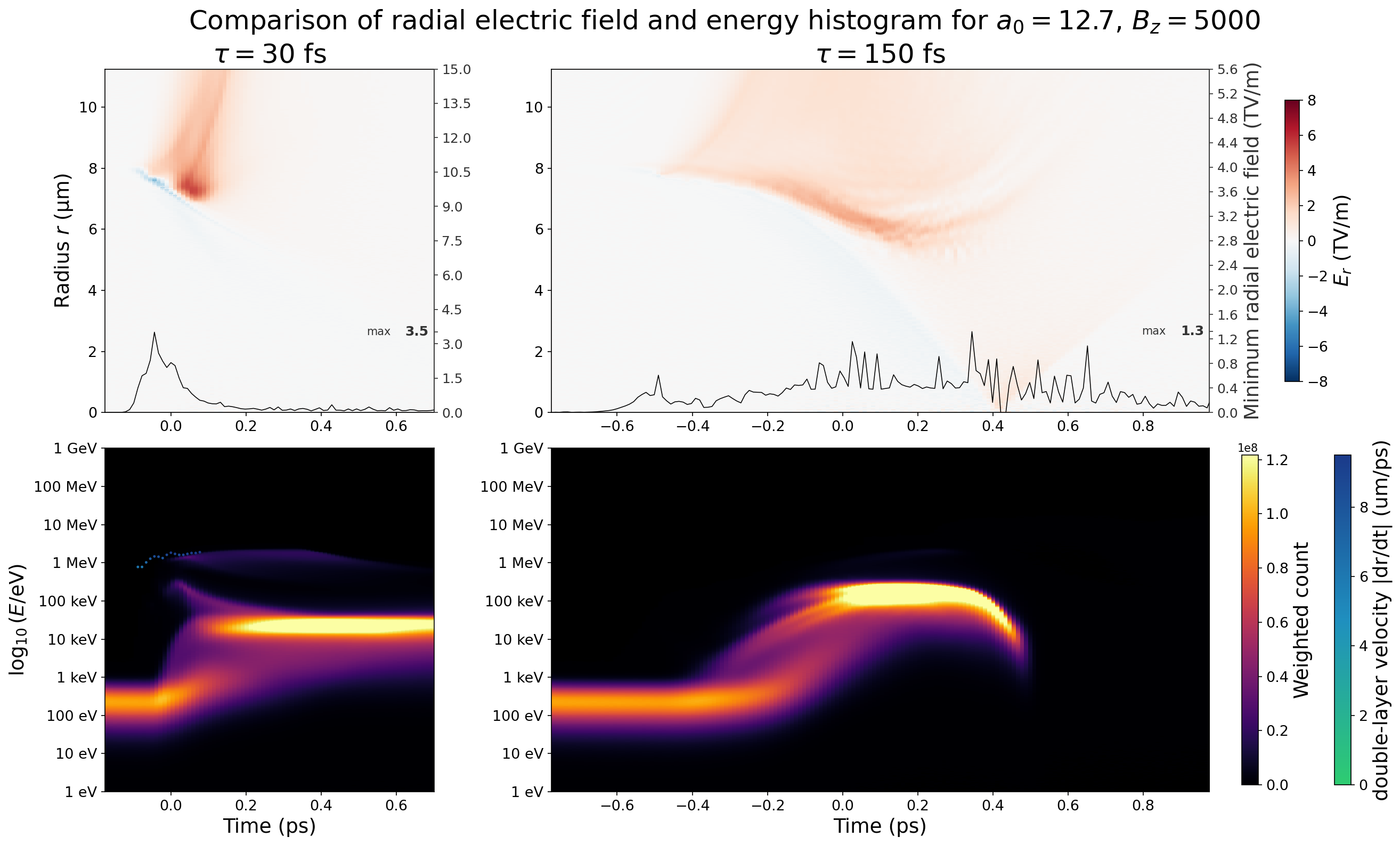}

  \vspace{0pt}

  \includegraphics[width=0.7\linewidth]{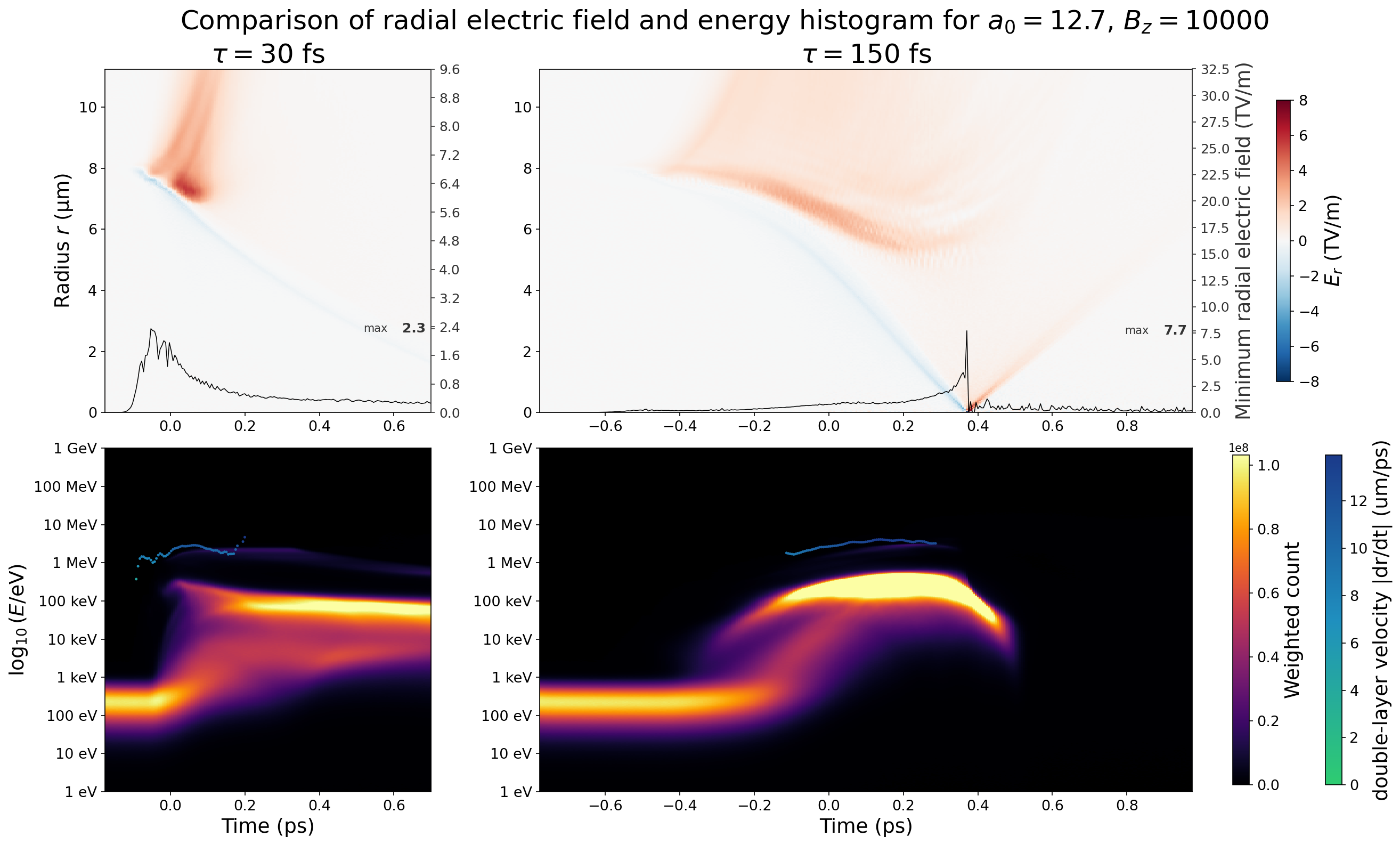}

  \caption{Suppression of the ion-accelerating charge-separation field with increasing axial magnetisation at fixed $a_0 = 12.7$. Each panel follows the layout of Figure~\ref{fig:kinematics22}: (top) $E_r(r,t)$ for the 30~fs and 150~fs drivers, with the annotated ``max'' giving the absolute peak of $|\min E_r|$ on the right-hand scale; (bottom) the inward-ion energy histograms with the $2v_{hb}$ reflection-energy markers overlaid (convention defined in the main text). Panels scan $B_z \in \{1, 5, 10\}$~kT. The $B_z = 0$ reference is shown in Figures~\ref{fig:kinematics12p7} and \ref{fig:kinematics22}.}
  \label{fig:Er_hist_Bz_sweep_a12}
\end{adjustwidth}
\end{figure}
\begin{figure}[htbp]
\begin{adjustwidth}{-\extralength}{0cm}
  \centering

  \includegraphics[width=0.7\linewidth]{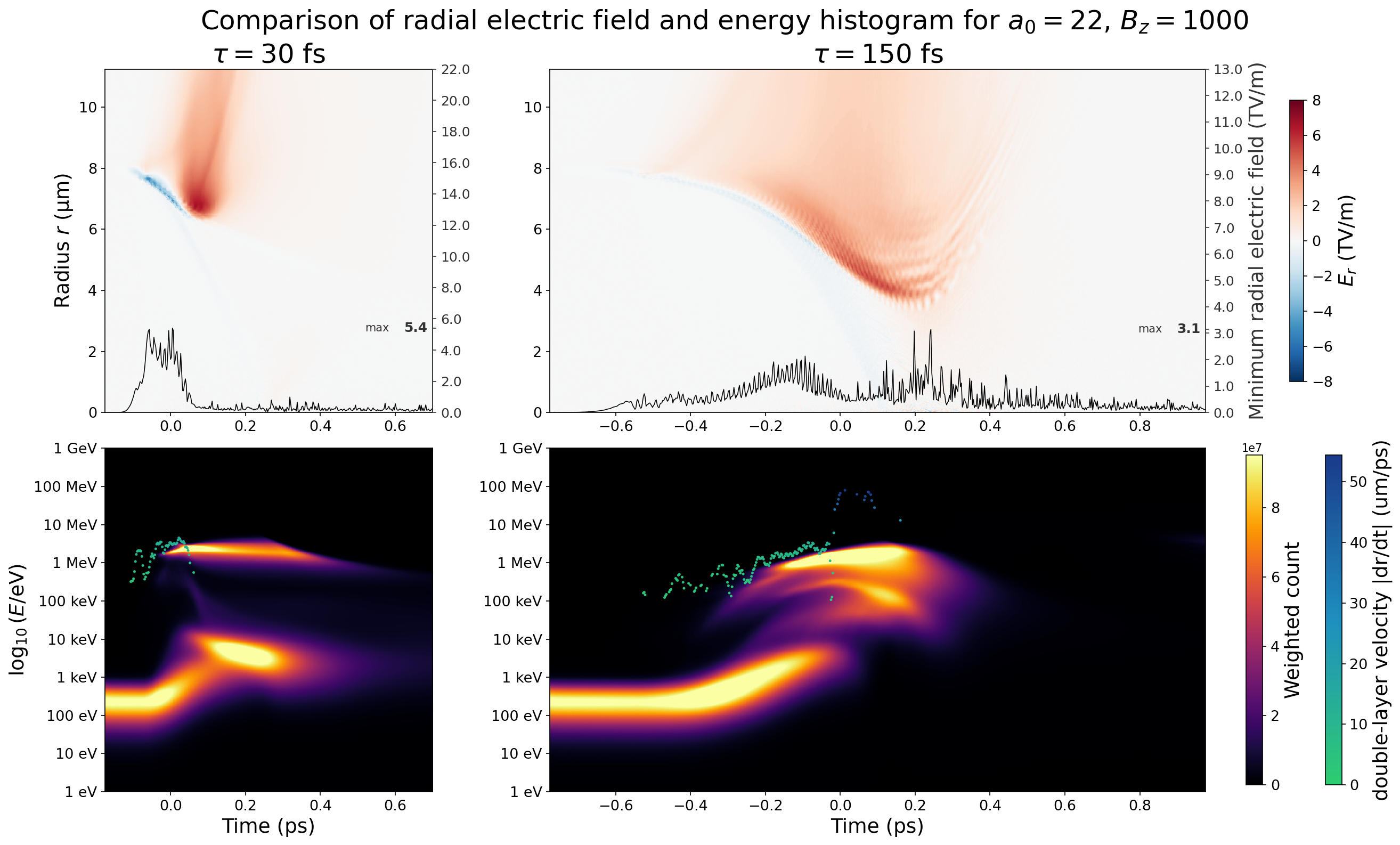}

  \vspace{0pt}

  \includegraphics[width=0.7\linewidth]{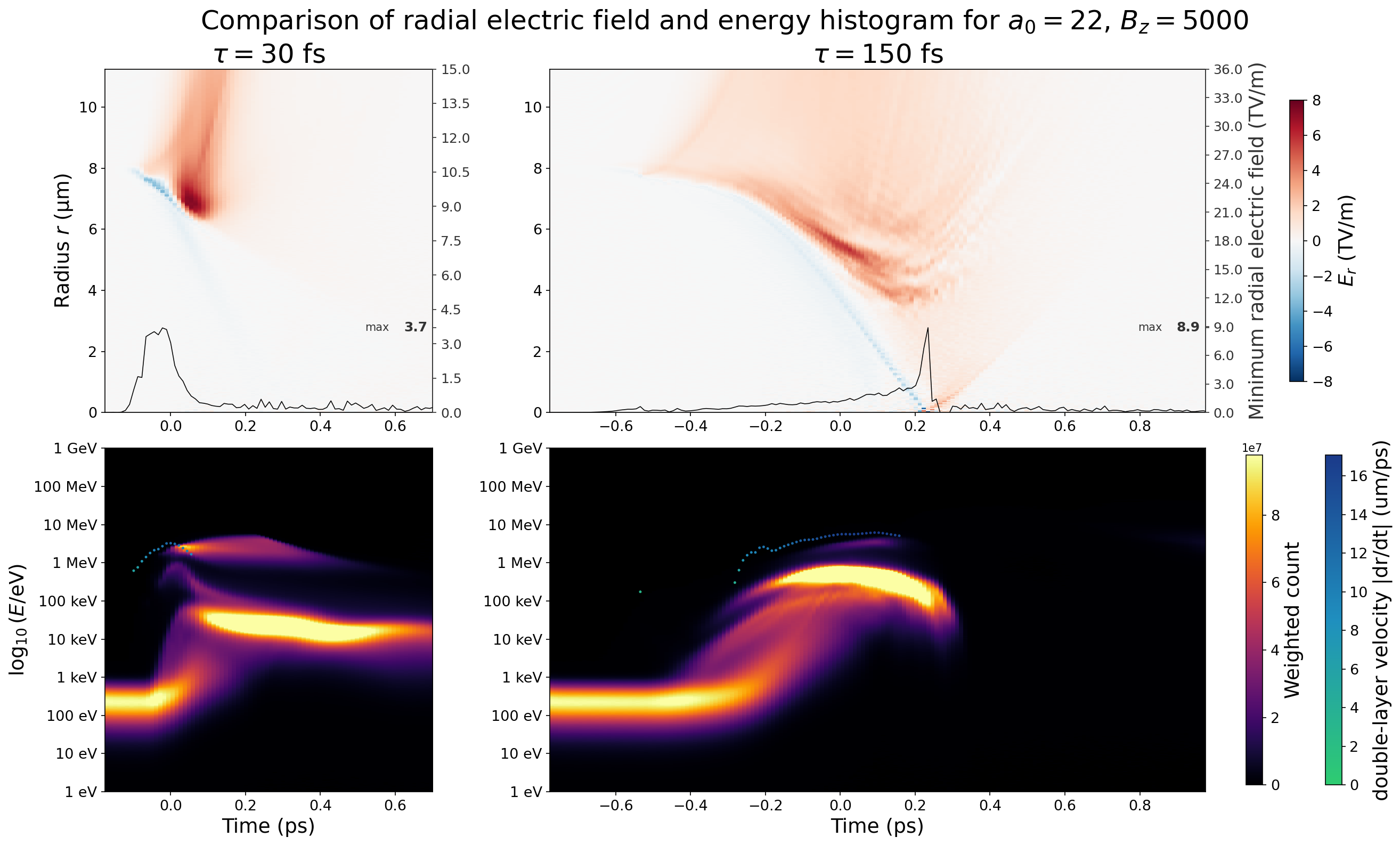}

  \vspace{0pt}

  \includegraphics[width=0.7\linewidth]{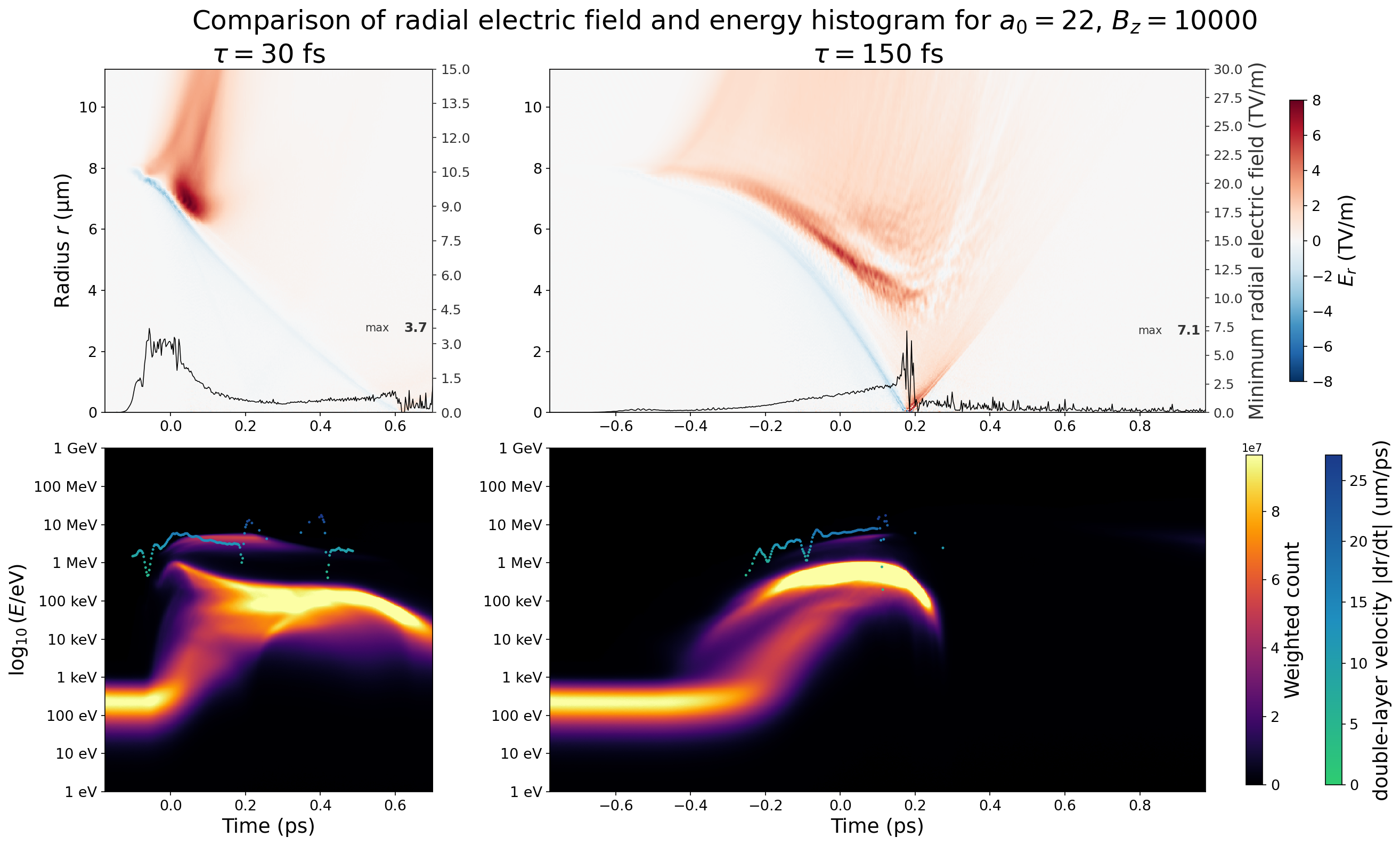}

  \caption{Same as Figure~\ref{fig:Er_hist_Bz_sweep_a12}, but for $a_0 = 22.0$.}
  \label{fig:Er_hist_Bz_sweep_a22}
\end{adjustwidth}
\end{figure}
\begin{figure}[htbp]
\begin{adjustwidth}{-\extralength}{0cm}
  \centering
  \includegraphics[width=0.85\linewidth]{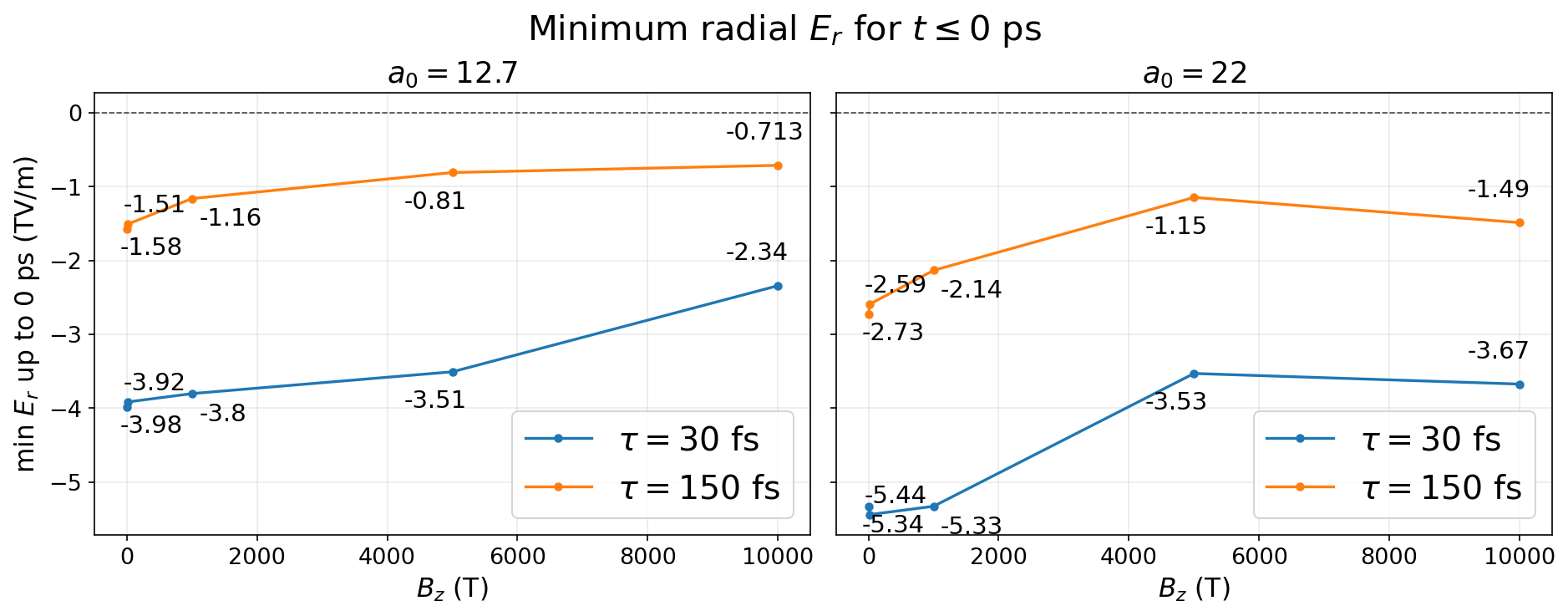}
  \caption{Single-scalar summary of the ion-accelerating field: $\min_{t \le 0}\, E_r$ as a function of $B_z$ for each of the four $(\tau, a_0)$ configurations. Left panel: $a_0 = 12.7$; right panel: $a_0 = 22.0$. Blue curves: $\tau = 30$~fs; orange curves: $\tau = 150$~fs. The restriction to $t \le 0$~ps (pre-peak phase at the target surface) excludes the transient spike produced when the double-layer collapses at $r = 0$, so that the plotted quantity genuinely reflects the inward-propagating, ion-accelerating front rather than the central-focusing artefact discussed in the main text.}
  \label{fig:min_Er_vs_Bz}
\end{adjustwidth}
\end{figure}

\begin{adjustwidth}{-\extralength}{0cm}

\subsection{Blowoff behaviour}\label{sec:blowoff}

A feature of the magnetised runs, visible directly in volumetric renderings of the ion kinetic-energy density, is that the outermost target envelope extends \emph{further} from the target centre as $B_z$ is increased, for otherwise identical laser parameters. The effect is visible for both species---electrons and ions. One possible contributing mechanism is that electrons that are pinned by the magnetic field can also experience ponderomotive expulsion on the downstream side of the laser field enhancing the outward envelope. Figure~\ref{fig:envelope_overlay} shows the comparison directly. For the 30~fs driver (left) the unmagnetized envelope (red) is the most compact, and the envelope grows monotonically outward through $B_z = 5$~kT (grey) to $B_z = 10$~kT (blue), with a clearly visible separation of the three outer hulls. For the 150~fs driver (right) the same ordering holds across all four overlaid fields ($0$~T red, $1$~kT blue, $5$~kT grey/white, $10$~kT golden), although the difference between the $5$~kT and $10$~kT envelopes is much smaller than for the 30~fs driver---consistent with the weaker overall sensitivity of the long-pulse case to magnetisation that we already saw in the all-species $t_{\text{comp}}$ of Figure~\ref{fig:t_comp_summary}.

\end{adjustwidth}
\begin{adjustwidth}{-\extralength}{0cm}

A naive picture in which the axial magnetic field simply confines the plasma---electron Larmor radii shrink to $\lesssim 1~\mu$m at $10$~kT (Table~\ref{tab:larmor}), and ions are not magnetised at all below $10$~kT---predicts the opposite ordering for electrons and no effect at all for ions. The data show neither. At the descriptive level, increasing $B_z$ makes the electron population less mobile across the target scale, so that the laser-driven electron response becomes less purely forward (inward): electrons can remain closer to the laser interaction region for longer and experience both the inward ponderomotive push on the rising edge and an outward expulsion after passing the intensity peak. The resulting time-extended charge separation can then contribute to a stronger late-time outward electric field and, through the ambipolar coupling between electrons and ions, a larger final ion envelope (Figure~\ref{fig:envelope_overlay}). A definitive separation of these effects is left for future work.

\subsection{Suppression of the Ion-Acceleration Mechanism with Increasing $B_z$}\label{sec:ion_suppression}

The magnetised-sheath picture of Section~\ref{sec:blowoff} makes a specific, falsifiable prediction for the ion-driving electric field at the target surface: the charge-separation field $E_r$ should weaken overall as $B_z$ grows (because the streaming hot electrons that sustain the laser-driven piston are increasingly locked onto field lines and no longer contribute to the front). We now revisit $E_r(r,t)$ and the inward-ion energy histogram---the two diagnostics that in Section~\ref{sec:magnetic_results}'s unmagnetized counterpart (Figures~\ref{fig:kinematics12p7} and \ref{fig:kinematics22}) established the charge-separation front as the driver of the fast-ion population---and repeat them across the full high-field sweep, $B_z \in \{1, 5, 10\}$~kT, at both intensities $a_0 = 12.7$ and $a_0 = 22.0$. The resulting six panel pairs are collected in Figures~\ref{fig:Er_hist_Bz_sweep_a12} and~\ref{fig:Er_hist_Bz_sweep_a22}; each panel internally contrasts the 30~fs and 150~fs drivers, exactly as in Figure~\ref{fig:kinematics22}.

\end{adjustwidth}
\begin{adjustwidth}{-\extralength}{0cm}

\end{adjustwidth}
\begin{adjustwidth}{-\extralength}{0cm}

The dominant trend across Figures~\ref{fig:Er_hist_Bz_sweep_a12} and~\ref{fig:Er_hist_Bz_sweep_a22} is a progressive weakening of the ion-accelerating field with increasing magnetisation: as $B_z$ is increased from 1~kT to 10~kT, the inward-pointing (blue) lobe of $E_r$ that marks the moving charge-separation front becomes weaker but longer-lived, and the MeV-class fast-ion band in the histograms progressively disappears. This is the direct mechanistic counterpart of the hot-electron trapping picture developed in Section~\ref{sec:magnetic_results} and quantified by Table~\ref{tab:larmor}: as the MeV-scale electron Larmor radius drops below the target size, the laser-driven piston that would otherwise displace electrons inward and open the charge-separation gap loses the electron flux needed to sustain it, and the ion-accelerating field is quenched. The effect is present at both $a_0 = 12.7$ and $a_0 = 22.0$; at fixed $B_z$ the higher-intensity case retains a larger residual field, consistent with its stronger ponderomotive drive, but follows the same qualitative suppression trend.

Note, the headline ``max'' annotation in Figures~\ref{fig:Er_hist_Bz_sweep_a12} and~\ref{fig:Er_hist_Bz_sweep_a22} reports the single largest value of $|\min E_r|$ attained at any time during the run, and this metric is not always representative of the acceleration-relevant field strength. For the 150~fs driver in particular, the sustained charge-separation front continues to focus as it converges on the target axis, so $|\min E_r|$ spikes transiently when the double-layer crosses $r = 0$; this central-collapse event has no bearing on the inward ion acceleration that took place during the preceding hundreds of femtoseconds but nevertheless inflates the ``max'' label. To obtain a single-number summary that is free of this artefact, we collapse each run to $\min_{t \le 0}\,E_r$, the most strongly inward-pointing radial field attained during the \emph{rising edge} of the pulse only. Restricting the time window to $t \le 0$~ps excises the central-collapse spike by construction.

Figure~\ref{fig:min_Er_vs_Bz} plots this quantity against $B_z$ for each of the four $(\tau, a_0)$ configurations, one panel per intensity, with one curve per pulse duration. The dominant trend is a systematic suppression of the inward field with increasing magnetisation: for every driver the amplitude decreases between $B_z = 0$ and $B_z = 5$~kT, and the 30~fs driver remains stronger than the 150~fs driver by a factor of two to three at every $B_z$---the hierarchy of drivers established in Section~\ref{sec:magnetic_results} for the unmagnetized case is preserved under magnetisation.

Two points deviate from this suppression pattern, both at $B_z = 10$~kT and $a_0 = 22.0$: the 30~fs value rises back up relative to $B_z = 5$~kT, and the 150~fs value shows a weaker version of the same reversal. This is a physical effect, not numerical scatter. At high axial fields the hot electrons are sufficiently magnetised (see Table~\ref{tab:larmor}) that the charge-separation double-layer no longer dissipates promptly when the ponderomotive drive turns off; the structure survives the end of the pulse and continues to converge toward the target axis under its own inertia. As the persistent layer sweeps inward, the cylindrical geometry concentrates the same separated charge into a progressively smaller radial shell, and the electric field grows as the shell volume shrinks---first linearly and then, as $r\to0$, far more steeply. The 30~fs / $a_0 = 22.0$ / $10$~kT run captures this late-time geometric focusing within the $t \le 0$ window used for the diagnostic, pushing its entry above the $5$~kT point; the 150~fs version of the effect is weaker because its longer rising edge already drives a more extended double-layer whose late-time convergence is smoother and less focused. At $a_0 = 12.7$ the ponderomotive drive is not sufficient for this residual layer to dominate the measurement, and the suppression remains monotonic across the full $B_z$ range.

\end{adjustwidth}
\begin{adjustwidth}{-\extralength}{0cm}

Taken together, the $E_r(r,t)$ maps of Figures~\ref{fig:Er_hist_Bz_sweep_a12} and~\ref{fig:Er_hist_Bz_sweep_a22}, the $\min_{t \le 0}\,E_r$ curves of Figure~\ref{fig:min_Er_vs_Bz}, and the Larmor-radius values of Table~\ref{tab:larmor} form a consistent picture. Magnetisation of the hot-electron population at $B_z \gtrsim 1$~kT progressively quenches the laser-driven charge-separation mechanism that, in the unmagnetized regime, produces the $\sim 1$--$5$~MeV fast-ion band---and, at the same time, causes the residual double-layer to persist long enough to eventually focus on the target axis, producing the single non-monotonic reversal observed at the strongest field and highest intensity. The ion-acceleration channel that dominates in the unmagnetized regime is therefore strongly suppressed; any residual late-time field amplification is geometric rather than laser-driven, and its implications for core energy deposition remain to be assessed in dedicated 3D follow-up studies.

\section{Discussion and Future Outlook}

\subsection*{Driver Comparison: Impulsive vs.\ Sustained Piston}
The two laser drivers studied here produce qualitatively distinct interaction regimes that persist across all magnetic field strengths and both intensities. The DRACO-class 30~fs pulse delivers an impulsive ponderomotive kick: the charge-separation front forms and dissipates within a few tens of femtoseconds, reflecting a small fraction of ions into a monoenergetic $\sim 1$~MeV band while leaving the bulk to undergo ablation-dominated decompression. The PEnELOPE-class 150~fs pulse, by contrast, acts as a sustained piston throughout its extended rising edge: the front continuously accelerates for several hundred femtoseconds, sweeping the fast-ion energy from hundreds of keV up to a few MeV and simultaneously driving the bulk to $\sim 80$~keV. These two regimes are not merely quantitatively different---they are mechanistically distinct, and both are captured with full fidelity only in a kinetic framework. The common thread is the $2v_{hb}$ reflection model: because the charge-separation front is intrinsically non-quasi-neutral (Section~\ref{sec:sound_speed_vs_dl}), the reflection kinematics are entirely set by the instantaneous front velocity, and a single scalar extracted from the $E_r(r,t)$ data quantitatively predicts the fast-ion energy at every time and for every $(\tau, a_0)$ combination studied.

\subsection*{Magnetisation Threshold and the kT-Scale Physics}
The magnetic-field scan reveals a sharp hierarchy of effects. The lab-achievable 20~T regime is quantitatively inert: every relevant particle species---MeV hot electrons and protons alike---has a Larmor radius exceeding the target diameter by at least one order of magnitude (Table~\ref{tab:larmor}), and no macroscopic observable is modified. The onset of target-scale magnetisation for the MeV hot-electron population requires fields of order 1~kT, above which a coherent set of signatures emerges simultaneously: the ion-accelerating charge-separation field is progressively quenched, the MeV fast-ion band disappears from the energy spectra, the net-inward compression time of the 30~fs driver more than doubles, and the peak core density develops a second, larger bulk-driven maximum that is absent in the unmagnetized case. The blowoff---that the outer target envelope grows with $B_z$ rather than shrinking---is consistent with this picture: magnetically pinned hot electrons sustain a late-time outward ambipolar field that expands the ion envelope even as the core is being compressed more effectively.

\subsection*{Experimental Accessibility and Geometric Equivalence}
The 1--10~kT fields necessary to trigger magnetised-sheath physics cannot be applied externally to a 15~$\mu$m uncompressed hydrogen target, and the lab-standard 20~T regime is macroscopically inert (Section~\ref{sec:20T}). Accessing this physics experimentally therefore requires one of two approaches. The first utilizes self-generated fields: tilted-incidence irradiation drives axial return currents yielding kT-class azimuthal fields ($B_\phi$) at the surface, though their coupling to radial compression remains an open question. The second exploits geometric equivalence. Because the governing dimensionless parameter is the ratio of the electron Larmor radius to the target radius ($r_L / R$), a 10~kT field on a 7.5~$\mu$m target is kinematically equivalent to a $\sim$10~T field on a 7.5~mm target. Therefore, the extreme confinement physics reported here maps directly onto larger, experimentally standard targets. Both of these non-mutually exclusive routes offer strong avenues for future study.

\subsection*{Mitigation of Azimuthal Asymmetries}
As visually evident in the kinetic energy density evolution (Figure~\ref{fig:kinetic_map}) and the final target envelope renderings (Figure~\ref{fig:envelope_overlay}), the three-beam irradiation drives a pronounced triangular, non-uniform compression front. This lack of azimuthal symmetry leaves ``gaps'' through which particles can escape the target. Future numerical setups could investigate enlarging the laser focal spot ($w_0 > 7.5$ $\mu$m) to increase beam overlap at the target surface. However, we note that expanding to a full 3D geometry---incorporating spherical divergence and longitudinal current topologies---may fundamentally alter these smoothing effects, a dynamic that remains to be explored in future 3D simulations.

\subsection*{Limitations of the Present Study}
The 2D3V geometry imposes two notable constraints. First, the out-of-plane direction is treated as a uniform slab, which suppresses longitudinal surface return currents; in a full 3D geometry these currents modify the current-closure topology and may alter both the self-generated field structure and the electron transport. Second, the magnetic mapping from kT-scale fields to larger targets described above is a kinematic equivalence---it preserves the Larmor-radius-to-target-radius ratio---but does not constitute exact plasma similarity, which would additionally require the density to scale as $1/R^2$. In particular, the absolute field amplitudes, peak compression ratios, electron-ion collision frequency $\nu_{ei}$ (and hence the Hall parameter $\omega_{ce}/\nu_{ei}$ that governs the magnetised-transport regime), and the absolute compression and gyration timescales do not transfer one-to-one to a larger-diameter target under this mapping; only the ratio of the electron Larmor radius to the target radius is preserved. The reported compression ratios and absolute field values should therefore be interpreted as indicative rather than directly scalable. Finally, the interactive data repository \cite{Optolowicz2026} makes all simulation outputs available for independent analysis.

\subsection*{Future Work}

Several natural extensions follow from the present results. Three-dimensional simulations are needed to capture longitudinal return currents and to resolve the full cylindrical symmetry of the three-beam geometry. Tilted-incidence runs with PIConGPU would allow us to study the self-generation of axial fields and their interplay with the radial compression, closing the loop between the applied-field scan presented here and experimentally realistic field configurations. On the target side, extending the simulations to deuterium-tritium fuel would enable direct tracking of the fusion-relevant nuclear reactions and would allow the neutron yield from laser-driven colliding beams to be estimated as a function of the driver parameters identified in this work. Finally, direct comparison with the experimental observables at DRACO and the future PEnELOPE facility---ion energy spectra, shadowgraphy of compression-wave convergence, and X-ray emission timing---could provide the experimental validation of the predictive baseline established here.


\vspace{6pt} 

\authorcontributions{Conceptualization, F.O., K.S. and D.B.; methodology, F.O. and B.M.; software, F.O. and B.M.; formal analysis, F.O.; investigation, F.O.; data curation, F.O.; writing---original draft preparation, F.O.; writing---review and editing, F.O., K.S., D.B., M.B. and B.M.; visualization, F.O.; supervision, K.S., D.B. and M.B.; project administration, K.S. and M.B.; funding acquisition, M.B. All authors have read and agreed to the published version of the manuscript.}

\funding{This project received access to the JUPITER supercomputer, which is funded by the EuroHPC Joint Undertaking, the German Federal Ministry of Research, Technology and Space, and the Ministry of Culture and Science of the German state of North Rhine-Westphalia, through the JUPITER Research and Early Access Program (JUREAP).
}

\acknowledgments{The authors acknowledge discussions with participants at the Hungarian-German WE-Heraeus Seminar on ``Particles and Plasmas in Strong Fields'', Dresden and G\"orlitz, June 22--26, 2025 and at the 46th International Workshop on ``High-\-Energy-Density Physics with Intense Ion and Laser Beams'', Hirschegg, January 25--30, 2026, where the preliminary results have been presented.}

\section*{Disclaimer}

This manuscript was prepared with the assistance of AI tools to support the writing and analysis workflow. Post-processing scripts were developed with GitHub Copilot and Google Gemini. The textual content was restructured, rephrased, and refined using Cursor (employing models from Anthropic, Google, and Cursor), Google Gemini (versions 3 pro and 3.1 pro), and Anthropic Claude (versions 4.5 and 4.6). These tools were used interactively rather than as passive formatters: in particular, Gemini often suggested directions for further investigation and helped organise the narrative, while the underlying physical concepts, simulation design, and interpretive choices remained driven by the authors.

The AI assistants did not supply new scientific results. All simulation data, figures, conclusions, and analyses presented here are the product of the authors' own work. The AI tools were used to improve clarity, structure, and exposition without altering the substance of the research or its findings.

\reftitle{References}


\end{adjustwidth}

\end{document}